\documentclass{aa}

\usepackage{graphicx}

\usepackage{xcolor}
\usepackage{txfonts}
\usepackage{hyperref}
\usepackage{subcaption}

\defcitealias{Brandt2017}{B17}

\begin{document}

   \title{Spectral cube extraction for the VLT/SPHERE IFS\thanks{The pipeline and documentation are publicly available at \url{https://github.com/PrincetonUniversity/charis-dep}.}}
   \subtitle{Open-source pipeline with full forward modeling and improved sensitivity}
   \author{M.~Samland\inst{1},
                   T.~D.~Brandt\inst{2},
                   J.~Milli\inst{3},
                   P.~Delorme\inst{3},
                   A.~Vigan\inst{4}}

\institute{Max-Planck-Institut f\"ur Astronomie, K\"onigstuhl 17, 69117 Heidelberg, Germany\\
\email{samland@mpia.de}
\and
Department of Physics, University of California, Santa Barbara, Santa Barbara, CA 93106, USA
\and
Univ. Grenoble Alpes, CNRS, IPAG, F-38000 Grenoble, France
\and
Aix Marseille Univ, CNRS, CNES, LAM, Marseille, France
}
   \date{Received 25 July 2022; accepted 16 September 2022}

  \abstract{We present a new open-source data-reduction pipeline to reconstruct spectral data cubes from raw SPHERE integral-field spectrograph (IFS) data.  The pipeline is written in Python and based on the pipeline that was developed for the CHARIS IFS. It introduces several improvements to SPHERE data analysis that ultimately produce significant improvements in postprocessing sensitivity.  We first used new data to measure SPHERE lenslet point spread functions (PSFs) at the four laser calibration wavelengths.  These lenslet PSFs enabled us to forward-model SPHERE data, to extract spectra using a least-squares fit, and to remove spectral crosstalk using the measured lenslet PSFs.  Our approach also reduces the number of required interpolations, both spectral and spatial, and can preserve the original hexagonal lenslet geometry in the SPHERE IFS. In the case of least-squares extraction, no interpolation of the data is performed.  We demonstrate this new pipeline on the directly imaged exoplanet 51~Eri~b and on observations of the hot white dwarf companion to HD~2133.  The extracted spectrum of HD~2133B matches theoretical models, demonstrating spectrophotometric calibration that is good to a few percent.  Postprocessing on two 51~Eri~b data sets demonstrates a median improvement in sensitivity of 80\% and 30\% for the 2015 and 2017 data, respectively, compared to the use of cubes reconstructed by the SPHERE Data Center. The largest improvements are seen for poorer observing conditions.  The new SPHERE pipeline takes less than three minutes to produce a data cube on a modern laptop, making it practical to reprocess all SPHERE IFS data.}


   \maketitle
%

\section{Introduction}
In recent years, a new generation of integral-field spectrographs (IFSs) on 8-meter class telescopes have been developed especially with the purpose of aiding the direct detection and characterization of extrasolar planets. Integral-field spectrographs allow us to obtain spectral information across the field of view (FoV) of the instrument, hence providing spatially resolved images at a wide range of wavelengths at the same time.

Integral-field spectrographs have proven to be a powerful asset in the toolbox of high-contrast imaging; not only can the spectral information be used to characterize the properties of any astrophysical object detected, it can also be used to exploit the different chromatic behavior of diffraction speckles from a real astrophysical source to suppress noise and improve detection contrast \citep[e.g.,][]{Racine1999, Sparks2002, Hoeijmakers2018}.

The recent generation of IFS instruments combines the high angular resolution of the telescope with extreme adaptive optics (XAO) to provide near diffraction-limited images of stars and their surroundings \citep[e.g.,][]{Guyon2005ApJ, Jovanovic2015b}. The difference in brightness between the star and potential planets necessitates the use of further technologies such as coronagraphs \citep[e.g.,][]{Soummer2005, Guyon2005AAS, Snik2012} to suppress the light from the central star. Dedicated high-contrast imaging techniques are also needed to distinguish between the unavoidable residual starlight halo and genuine astrophysical signals in the direct vicinity of the star, such as angular differential imaging \citep[ADI;][]{Marois2006} and spectral differential imaging \citep[SDI;][]{Racine1999}.

Currently, four such dedicated (XAO) high-contrast imaging instruments are equipped with an IFS: GPI on Gemini South \citep[soon Gemini North,][]{Macintosh2014}, ALES at the Large Binocular Telescope \citep[LBT;][]{Skemer2015}, SPHERE at the Very Large Telescope \citep[VLT;][]{Beuzit2019}, and CHARIS on the Subaru telescope \citep{Groff2015}. All of these instruments use a similar technology that is based on lenslet arrays to map spatial elements onto small spatially separated spots that will be dispersed onto the detector as distinct spectra. Although these instruments differ in implementation details and design (e.g., wavelength range, resolution, and size of the FoV), they are sufficiently similar for a suitably flexible software package to be adapted to the peculiarities of each individual instrument.

The reduction and processing of IFS data is inherently complex. After investing a large part of their budgets in the initial development and building of an instrument, individual teams therefore often struggle to provide the manpower for continuous support and improvement of the software that is used for data reduction. This can be remedied by providing the scientific community access to source codes and encourage community involvement in their further development and in improving their reliability.
In this paper we present our effort to take the first step in this direction by adapting the CHARIS data-reduction pipeline \citep[][henceforth B17]{Brandt2017}, an open-source package written in Python and Cython available on Github, to be able to process SPHERE IFS data into scientifically usable products. The largest difference of the SPHERE IFS compared to the other IFS instruments is its hexagonal lenslet geometry, known as BIGRE \citep{Antichi2009}. This will receive special attention in this work because this fact is not widely known to end-users of the data and has implications for future improvements of data-reduction algorithms and planet detection algorithms.

The pipeline described in this paper offers various improvements over the original SPHERE pipeline in how the spectral extraction is performed, and it addresses known shortcomings. 
The paper is structured as follows. Section~\ref{sec:sphere} gives an overview of the design properties and peculiarities of the SPHERE IFS. It also describes the current ESO SPHERE pipeline. Section~\ref{sec:calibration} shows the detailed steps we performed to obtain the necessary calibration data, and the static calibration products that are incorporated into the pipeline. Section~\ref{sec:extraction} describes the spectral extraction methods we implemented, the process that transforms raw data into the spectral image cubes. Section~\ref{sec:results} compares the results obtained with our new pipeline to results obtained with the official pipeline. Section~\ref{sec:usage_performance} details what the active user of the software needs to know to use the pipeline, and it provides a brief overview of the computational performance. Finally, we discuss our findings and conclude in section~\ref{sec:discussion_conclusion}.

\section{Overview of the SPHERE IFS and ESO pipeline}
\label{sec:sphere}
\begin{table}[t]
\caption[Basic properties of the SPHERE IFS]{Basic properties of the SPHERE IFS}
\centering
\begin{tabular}{l c}
\hline\hline
Parameter & SPHERE-IFS \\
\hline
Detector & $2048\times 2048$ H2RG \\
No. of lenslets & $\approx$1$52\times 152$\\
Lenslet lattice configuration & hexagonal \\
Lenslet area & 370 mas$^2$ \tablefootmark{,a}\\
Field of view& $1.\!\!''73 \times 1.\!\!''73$ \\
Wavelength coverage & 0.95 -- 1.66 $\muup$m\\
Microspectrum length & $\sim$39 pixel \tablefootmark{b}\\
Separation between spectra & $\sim$5 pixel \tablefootmark{b}\\
R = $\lambda$/$\delta\lambda$ & $\sim$55; $\sim$35 \tablefootmark{b}\\
Available modes & Y--J; Y--H\\
\hline
\end{tabular}
\label{tab:ifs_properties}
\tablefoot{
\footnotesize
\tablefoottext{a}{Area of hexagonal lenslet on sky. Resampled by the pipeline to $\sim$(7.4 mas)$^2$ per square pixel. The side length $t$ of the hexagon is 7.4 mas, and the area $A=3\sqrt{3}\,t^2/2$.}
\tablefoottext{b}{Microspectrum length, spacing, and R may vary slightly depending on definitions and may vary slightly across the FoV.}
}
\end{table}

The SPHERE IFS is part of the SPHERE instrument \citep{Beuzit2019}, which is located on the stable Nasmyth platform of the VLT, which is operated by the European Southern Observatory (ESO) as part of a system of three dedicated high-contrast imaging instruments: ZIMPOL \citep[visible light polarimetry;][]{Schmid2018}, IRDIS \citep[dual-band imager;][]{Dohlen2008}, and IFS \citep[ ][]{Claudi2008, Mesa2015}. These instruments are located behind a common path and infrastructure (CPI) module that feeds the subsystems with a highly stabilized, AO-corrected beam. The SAXO XAO system for SPHERE achieves Strehl ratios better than 90\% \citep{Petit2012} in the near-infrared (NIR). 
In this paper we focus exclusively on the extraction and analysis of data cubes from the IFS subsystem.
Table~\ref{tab:ifs_properties} lists the basic properties of the SPHERE IFS.

The SPHERE IFS has been used to discover and characterize exoplanets and brown dwarfs \citep[e.g.,][]{Chauvin2017_HIP65426, Keppler2018, Samland2017, Milli2017_HD206893, Chauvin2018_HD95086, Mueller2018, Cheetham2019, Maire2020_HD72946}.
It can also be used to study extended objects such as debris disks \citep[e.g.,][]{Boccaletti2015, Milli2017} and asteroids \citep{Hanus2017}.
Several large surveys have been performed using the instrument \citep[e.g.,][]{Chauvin2017, Bohn2020, Janson2021a, Bonavita2022}, and a considerable archive of hundreds of observation sequences exists. It is being used to study the demographics of young giant planets in the SHINE survey \citep[e.g.,][]{Vigan2021_shine}.

The SPHERE IFS images a $1.\!\!''73 \times 1.\!\!''73$ patch of sky onto a lenslet array.  Each hexagonal lenslet focuses the light incident on it to a point in the plane of a detector.  A prism residing between the lenslet array and the detector disperses each point into an about 39-pixel-long low-resolution spectrum.  These spectra may cover either the $Y$--$J$ spectral range at a spectral resolution $R = \lambda/\delta\lambda \approx 55$ (henceforth referred to as YJ-mode) or the $Y$--$H$ spectral range at a lower $R \approx 35$ (henceforth referred to as YH-mode). \footnote{The official designations of the modes are IRDIFS and IRDIFS\_EXT. In these modes, IRDIS dual-band imaging data are taken at longer wavelengths at the same time as the IFS data. As IRDIS is not the topic of this paper, we use a more intuitive designation for the IFS modes: YJ- and YH-mode.}

The SPHERE IFS has an internal calibration unit.  This unit can uniformly illuminate the detector to create a detector flat-field, or it can uniformly illuminate the lenslet array to create a lenslet flat.  The lenslet flat appears on the detector as a grid of dispersed microspectra.  The calibration unit also takes wavelength calibration data by illuminating the lenslets with lasers of known wavelength.  All of these data are taken during the day, both for calibration purposes and to monitor the health of the instrument. Dark and background images are likewise obtained by closing shutters at different positions (in front of the IFS module for darks, or in front of the SPHERE entrance shutter for instrument backgrounds) during daytime. This saves observing time during the night for actual science data. The calibration unit consists of a) a series of lamps, both narrow and broad band, for taking detector flat-fields and flat-fields taken through the integral field unit (IFU), and b) a series of monochromatic lasers for wavelength calibration.  For SPHERE, the `wavecal' calibration template that is routinely executed takes images with three (YJ-mode) or four (YH-mode) lasers simultaneously (see left panel, Fig.~\ref{fig:laser_flats}).

The first step in the analysis of SPHERE IFS data is to take the 2D array of microspectra and reconstruct a 3D data cube: an image at each sampled wavelength.  Subsequent analysis steps, usually referred to as postprocessing, subtract starlight to search for faint exoplanets, calibrate and extract spectra, and calculate contrasts and sensitivities.  The ESO \textit{\textup{data-reduction and handling}} (DRH) pipeline \citep{Pavlov2008} is designed to perform all steps that are required to reconstruct data cubes from SPHERE IFS raw data.  This includes everything from thermal background subtraction and flat-field correction to the more intricate steps of wavelength calibration and extraction of the spectra into usable spectral image cubes. The SPHERE Data Center\footnote{The SPHERE DC performs data reduction on request and also processes all SPHERE public data to make them available publicly. More information is available at \href{https://sphere.osug.fr/spip.php?rubrique16}{https://sphere.osug.fr/spip.php?rubrique16}} \citep[DC,][]{SPHEREDC}, as well as a third-party Python wrapper \citep{Vigan2020_pipeline}, both combine the DRH with additional functions and improvements.  These include a more efficient correction for bad pixels, a correction for spectral crosstalk caused by overlapping lenslet spectra, an additional step to correct erroneous wavelength calibration, and a correction for anamorphism in the instrument.

The DRH pipeline begins by subtracting backgrounds and flat-fielding the detector.  Backgrounds are significant because most of the instrument is not cryogenically cooled.  The DRH then uses the dispersed lenslet flat-field image to identify the illuminated pixels of all microspectra and creates a mask for all of them in a step called specpos. The information about the position of each lenslet and the pixels associated with each spectrum is used in further steps. The DRH then fits for a wavelength solution (wavecal), a mapping between lenslet, wavelength, and position on the detector, using images taken with lasers of known wavelength.  This mapping is required in order to extract the spectrum of each lenslet. The last calibration required by the DRH pipeline is a map of relative lenslet transmissions, called the instrument flat-field. This is measured using a spectral image cube extracted from a uniform illumination of the lenslet array and is used to correct the flux of each lenslet in the spectral extraction routine.%
The DRH pipeline uses aperture photometry to extract the lenslet spectra.  In this approach, the flux is summed across a predefined aperture in the direction perpendicular to the dispersion.  The total flux in this direction is then assigned to the wavelength of this pixel row of the spectrum.  Aperture photometry involves tradeoffs.  A larger aperture includes more of the flux, but adds additional read noise and photon noise from the background.  Because of the relatively close spacing of the microspectra in SPHERE, a larger aperture also includes more flux from neighboring spectra at other wavelengths (spectral crosstalk).  These shortcomings may be overcome by optimal extraction \citep{Horne1986}, as long as the line-spread function perpendicular to the dispersion direction (the instrumental profile) is known.  This has not previously been measured for SPHERE, preventing an implementation of more sophisticated spectral extraction approaches.%
The SPHERE instrument uses hexagonal lenslets that are arranged in a honeycomb pattern (for an example of this, see Fig.~\ref{fig:extraction_methods}).  The DRH resamples the microspectra onto a rectilinear grid (see also Fig.~\ref{fig:comparison1}).  This requires an interpolation and sacrifices some of the information in the raw data.  Previous generations of high-contrast instruments lacked spectral information and took data on rectilinear grids (the arrangement of pixels on nearly all detectors).  The interpolation step taken by the DRH enables established postprocessing routines to work on SPHERE data without accounting for a new geometry.
The use of aperture photometry of the SPHERE DRH to extract microspectra and its resampling of the data cubes onto a rectilinear grid suggest that a new extraction approach could result in higher-quality data cubes.  \citet{Berdeu+Soulez+Ferreol+etal_2020} suggested an alternative regularized least-squares approach to reduce SPHERE IFS data, called PIC, and saw a reduction in data artifacts and improvements in final postprocessed contrast.  In the following sections, we describe an independent data-reduction pipeline based on that of the CHARIS IFS \citep{Groff2016,Brandt2017}.  This approach revisits every aspect of the data reduction and also takes advantage of new calibration data.

\section{Calibrations}
\label{sec:calibration}
Calibration data for the SPHERE IFS consist of uniformly illuminated images of the detector for flat-fielding, dark images for background subtraction, and wavelength calibration frames.  The latter consist of the lenslet array uniformly illuminated by three or four lasers of known wavelength fed through an integrating sphere.  Dark images have significant counts because most of SPHERE is not cryogenic.  

Darks in SPHERE IFS images consist of two components: an undispersed background from scattered light and emission after the lenslet array, and a dispersed background that is imaged by the lenslet array into microspectra.  Sky images include both backgrounds, while dark images include only the undispersed background.  Our approach uses only the dark (thermal instrument background) images and the detector flat-fields.  From these, we aim to construct a detector flat-field, a bad-pixel mask, and an undispersed background template. All calibration files necessary to reduce SPHERE IFS data are packaged with our pipeline, except for the laser flat-field needed to adjust the wavelength solution for each observing night. We include these master calibrations rather than adopting nightly calibrations for two reasons.  First, each individual calibration contains its own realization of read noise and photon noise.  If nightly calibration frames are used, the same realization of noise is added to each science frame, potentially doubling the read and background noise.  Master calibrations avoid this drawback, thereby providing better results as long as the calibrations themselves are stable.  In the following sections, we demonstrate the requisite stability of both the background and the flat-fields.  Our second reason for supplying master calibrations is that calibration frames occasionally fail, for instance, due to an incorrect shutter position or an uneven illumination of the detector.  The use of curated master frames avoids the possibility of bad darks and/or flats from user error or hardware failure.

\begin{figure}
  \centering
  \includegraphics[width=\columnwidth]{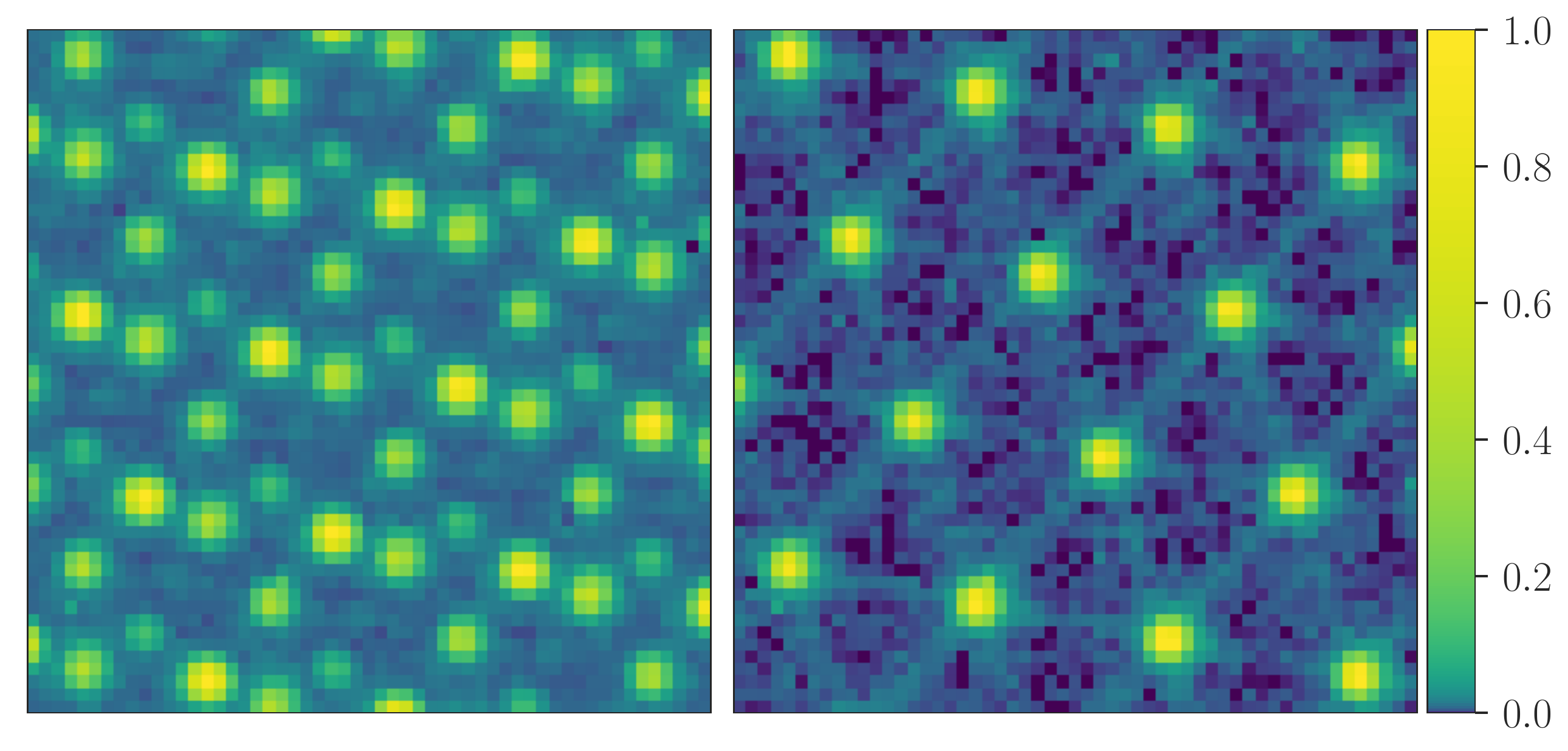}
  \caption{Normalized monochromatic laser flat-fields in logarithmic scale. The dispersion direction is from bottom to top. Left: Standard wavelength calibration image for the extended (YH) mode using all four lasers simultaneously. This is provided by ESO as part of the standard calibration sequence and is used by the ESO pipeline. Right: Similar image, but with only the 1123.7 nm laser on.  Images at this and three other wavelengths were taken and used to construct our lenslet PSF library (Section \ref{subsec:psflets}).}
  \label{fig:laser_flats}
\end{figure}

For this work, we have taken additional SPHERE IFS calibration data: images with only one laser illuminating the lenslets at a time.  This enabled us to clearly resolve the lenslet point spread functions (PSFs) and to extract the oversampled lenslet PSF in Section.~\ref{subsec:psflets}.  For YJ-mode, the laser wavelengths are 987.7~nm, 1123.7~nm, and 1309.4~nm. In YH-mode, we have one additional laser at 1545.1~nm.  Figure~\ref{fig:laser_flats} shows the difference between using all four lasers at the same time in YH-mode and using an individual laser. Both calibration images have been dark and flat corrected. 

\subsection{Bad-pixel mask}
Unfortunately, the detector in the SPHERE IFS has a significant number of bad pixels in clusters scattered throughout its FoV comprising $\approx$4.8\% of all pixels. The number of bad pixels further significantly increased after the pandemic shutdown in April 2020, during which the instrument was warmed up, to $\approx$5.0\%. Two master bad-pixel masks and master flat-fields for before and after the shutdown are therefore included with the pipeline to reflect the changes in the detector, which is otherwise very stable over time.

The first step of our new data-processing approach is to identify and construct a binary bad-pixel mask, which we did using the dark images and the detector flat-field images.  We first constructed a master dark from 100 (pre-shutdown) and 50 (post-shutdown) randomly selected dark calibration files, each containing ten exposures at an exposure time of 11 seconds. The master dark frames are a median combination of these 1000 or 500 dark frames, respectively. We then masked all pixels whose values exceed a 3.5$\sigma$ deviation in an iterative $\textrm{FWHM}=4$~pixel median filter kernel.  Finally, we constructed a master flat-field as described below in detail. We also masked pixels that differed significantly from the average value in a flat-field, along with pixels whose values fluctuated across flat-field images.  To do this, we masked all pixels that are below 90\% or more than 110\% of the median pixel counts or have a temporal standard deviation larger than 3\%.

Our approach identified a total of 167386 bad pixels from the dark frames and 31639 bad pixels from the flat-field images pre-shutdown, with numbers increasing by several thousand pixels post-shutdown.  We took the combined dark and flat mask as our final bad-pixel mask in both cases.
Our pipeline does not perform a bad-pixel interpolation because both spectral extraction methods, the optimal extraction (see Section~\ref{sec:optimal_extraction}) and the least-squares extraction approaches (see Section~\ref{sec:least_squares_extraction}), make it easy to exclude all bad pixels from the extraction process altogether. This means that interpolated bad-pixel data will not bias uncertainties due to newly introduced correlations. 
The SPHERE DRH pipeline, however, interpolates bad pixels based on all their neighbors, which is not ideal in the case of microspectra extraction.
For this reason, the SPHERE Data Center (DC) reductions \citep{SPHEREDC}, and the \texttt{vlt-sphere} Python pipeline \citep{Vigan2020_pipeline}, which implements a wrapper for the SPHERE DRH ESO pipeline and adds additional improvements, both implement an approach that linearly interpolates bad pixels using only the nearest good neighbor in the dispersion direction.

\subsection{Detector flat-field}
\label{sec:detectorflat}

The SPHERE IFS contains an internal calibration lamp that can uniformly illuminate the detector to correct for pixel-to-pixel variations in sensitivity.  We constructed the detector flat-field using two different exposure time settings. We randomly selected pairs of files of short (3s) and long (11s) exposures taken within one hour of each other (100 pairs before the pandemic shutdown; 50 pairs after the pandemic shutdown). Each file contains ten exposures, which we median combined. We took the difference image between corresponding long and short exposures to reduce nonsensitivity related effects and subsequently normalized them by their median while excluding already identified bad pixels. We discarded 17 visibly discrepant flat-fields pre-shutdown (none post-shutdown) that had a very uneven illumination of the detector, for example.  We performed no further pre-selection of frames. The remaining frames were median combined to obtain the respective master flat. 
We distribute the two flat-fields (before and after the shutdown) with the pipeline. The detector flat-field is stable with time at a level $\lesssim$0.35\%.  We do not divide our image by this detector flat, but rather use the flat later when extracting spectra: we multiply the PSFlet model counts by the flat-field rather than dividing the data by the flat-field. 

\begin{figure*}
\includegraphics[width=0.49 \textwidth]{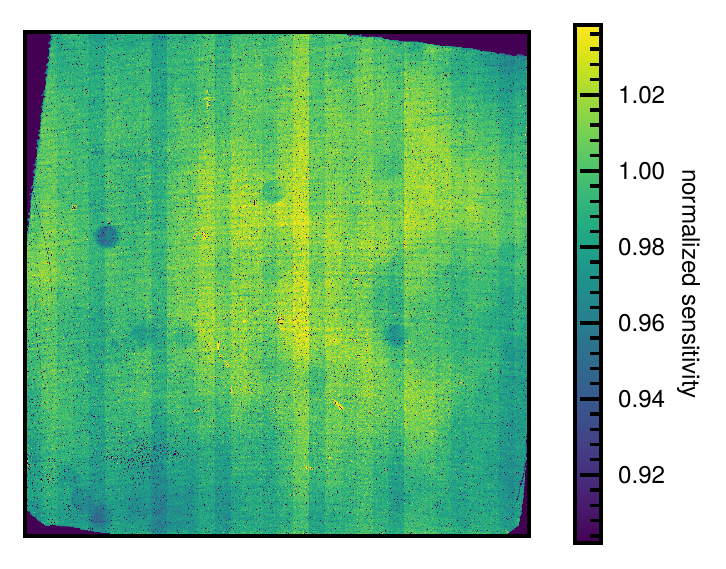}
\includegraphics[width=0.49 \textwidth]{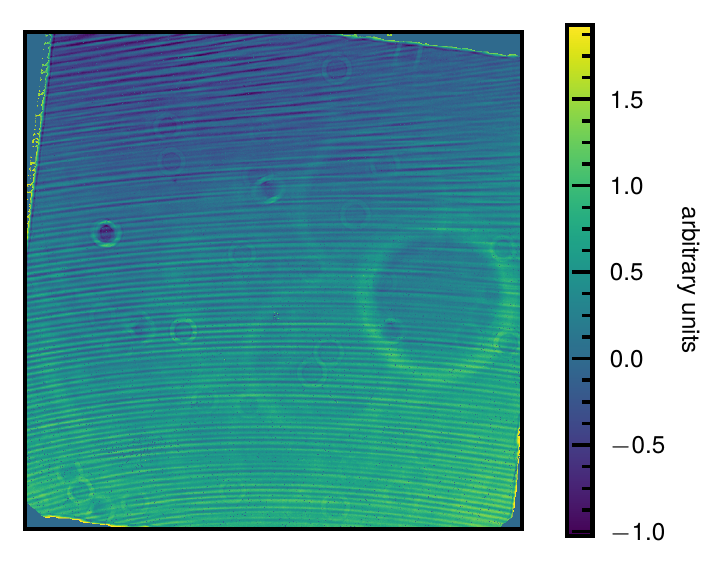}
\caption[IFS master flat-field and variable structures]{Overview of structures visible in the detector flat-field.} Left: Master flat-field generated by averaging hundreds of individual flat-field frames. Right: One component of an unranked empirical principal component decomposition of the temporal variance of the flat-fields into five basis vectors. This component highlights minute details of the flat-field variability. Details that  are normally not clearly visible become apparent due to the changes in illumination by the internal calibration lamp affecting diffraction patterns over time. This includes the fringe-like horizontal stripes, as well as dust diffraction by various optical surfaces.
\label{fig:flat_fields}
\end{figure*}

The left panel of Fig.~\ref{fig:flat_fields} shows the master detector flat-field distributed with our pipeline.  The flat is normalized to a median intensity of one.  The flat-field is not uniform, but it shows little variation with time.  We characterized this variation using principal component analysis (PCA): the right panel of Figure \ref{fig:flat_fields} shows one principal component of an unranked empirical PCA decomposition \citep{Bailey2012} of thousands of individual flat-field images taken at different times. This image highlights the structures that normally are not visible in individual exposures, providing a visual overview of the existing substructures in the flat, such as small dust on optical elements. 
The vertical stripes show the slightly different sensitivity of different readout channels of the H2RG detector.

The lenslet flat-field corrects for the variable transmission of lenslets in the lenslet array.  This requires a different approach from the detector flat-fielding described in this section.  Instead, we used images with SPHERE illuminated by a uniform white-light source (integrating sphere).  This produces microspectra that must be extracted to produce the lenslet flat-field.  We therefore defer discussion of the lenslet flat-field to Section~\ref{subsec:lenslet_flat} after we detail our approach to spectral extraction.  

\subsection{Undispersed background template}
\label{sec:background_template}
Because the SPHERE IFS is not cryogenic (except for the detector), background levels are high.  Much of this background is undispersed and illuminates the detector relatively uniformly.  We wish to remove this background before extracting the spectrum of each lenslet. All frames obtained with SPHERE are already corrected for additive bias by the read-out electronics.

Standard SPHERE observing sequences take a small number of sky frames after the science frames.  These sky frames may be used to remove both the undispersed background and the dispersed background (from the sky itself).  Unfortunately, using a sky frame for this purpose adds its noise to the science frame.  If both the sky frame and the science frame are of the same exposure time, this step increases the minimum photon noise by a factor of $\sqrt{2}$. Sky frames taken after the coronagraphic exposure also display noticeable persistence, which introduces additional noise.

In an effort to overcome this limit, we have constructed a master template for the undispersed background.  We seek a background without any dispersed light (e.g.,~from the sky), therefore we began with internal instrument background frames that do not contain sky.  We took 234 randomly selected background frames taken between 2015 and 2021 and performed PCA on them.  The number of frames was limited by the 64 GB of RAM on our computer and the need to perform PCA.  These frames come with the instrument in YH-mode. We repeated the exercise with YJ backgrounds and found that YH-derived calibrations fit YJ data just as well as YJ-derived calibrations do.
We used about 98\% of the pixels to perform PCA, masking the 1\% with the highest variance of the background frames and those with less than half or more than twice the average background level.

A mean plus three principal components offers excellent performance and accounts for slightly more than half of the mean-subtracted variance in the background frames. Additional principal components offer little improvement.  We fit a linear combination of these four images to the four corner regions of the detector that are unilluminated by lenslets.  In this way, we can fit our background templates to science frames and sky frames and preserve any light that is dispersed by the lenslet array and avoid the addition of read noise and photon noise.

\begin{figure}
    \centering
    \includegraphics[width=\linewidth]{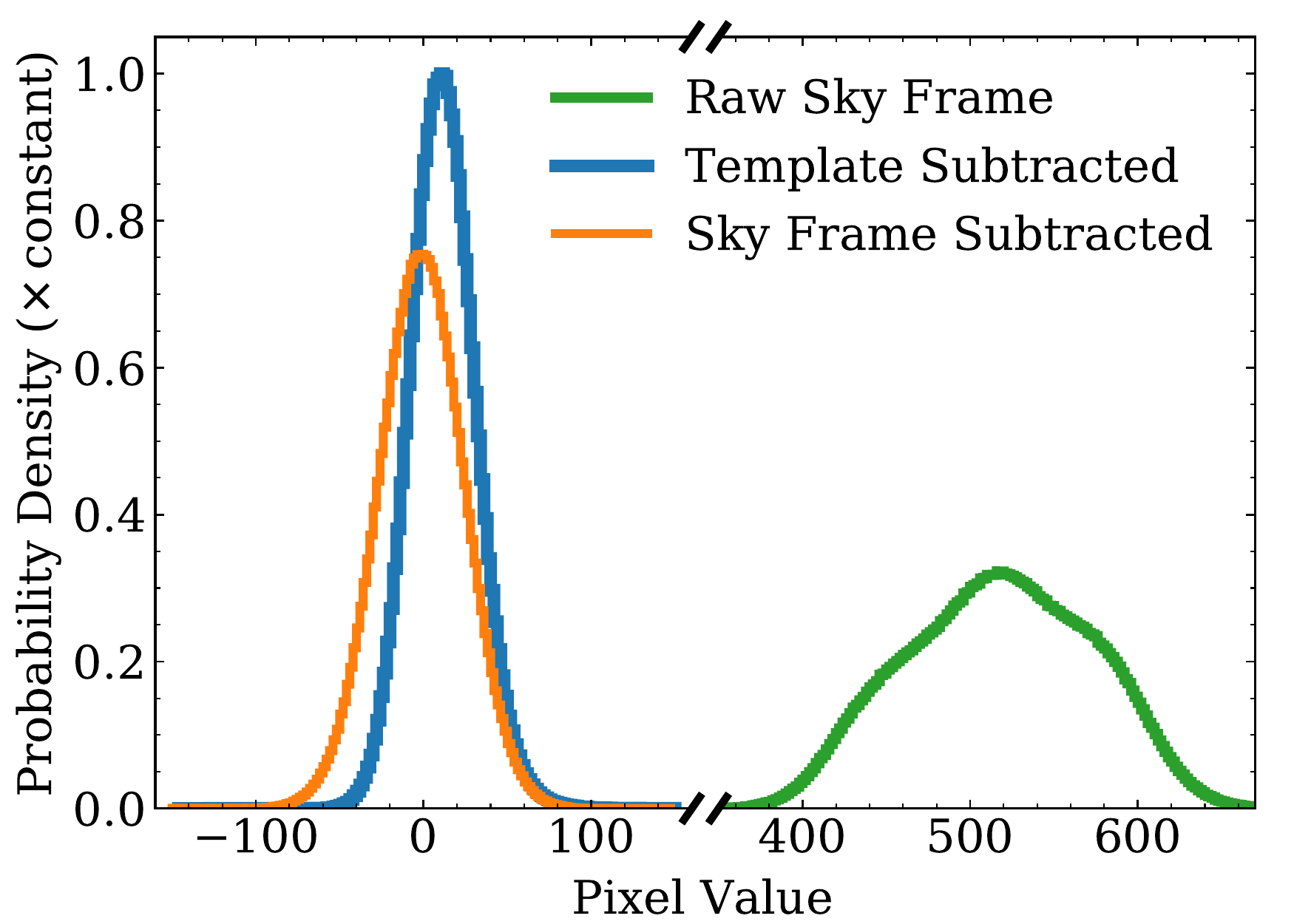}
    \caption{Distribution of pixel values in a raw 64-second sky frame (medium green histogram) after subtraction of another sky frame (thin orange histogram) and after subtraction of a background template (thick blue line).  }
    \label{fig:background_sub}
\end{figure}

Figure~\ref{fig:background_sub} shows the distribution of pixel values in a 64-second sky frame before and after background subtraction.  Subtracting another sky frame produces residuals that are centered about zero (the thin orange histogram) but adds read noise and photon noise. A different sky exposure may also contain varying levels of persistence, based on how brightly the coronagraphic images were exposed, which adds additional noise.  Subtracting the best-fit four component template produces a narrower distribution (thick blue histogram) because a negligible amount of additional noise is added.  This distribution is centered on a positive pixel value because it does not subtract the dispersed sky background and contains detector persistence that is intentionally not included in our model.  

Our SPHERE IFS pipeline implements the template subtraction described here for the undispersed background.  This approach minimizes the noise in the resulting data cubes.  A dispersed sky background does appear as a constant, uniform background in the extracted data cubes.  However, this is typically indistinguishable from the halo of the stellar PSF and is easily removed by postprocessing algorithms. Only when a library of PSFs is used that were taken on different nights might this introduce a small bias for extended sources if the dispersed background is significantly different. In this case, a constant may be fit to model the background.
Constructing a reference library of sky background frames is not feasible, as these are generally taken after the coronagraphic observations and show an imprint of the stellar halo due to detector persistence. This is another general issue with the normal sky subtraction approach that is typically performed, which adds a constant oversubtraction bias to all frames.

\subsection{Oversampled PSFlet templates} \label{subsec:psflets}

Spectral extraction requires at least a wavelength solution: a mapping between wavelength, lenslet, and location on the detector.  Many extraction techniques also require knowledge of the monochromatic PSF of each lenslet.  Optimal extraction requires the profile of this PSF (integrated along the dispersion direction), while least-squares extraction techniques require the full lenslet PSF henceforth, also referred to as PSFlet for convenience.  The standard SPHERE calibration images include either three or four wavelengths simultaneously, creating an image like that shown in the left panel of Fig.~\ref{fig:laser_flats}.  The simultaneous presence of multiple wavelengths is not a major problem for deriving a wavelength solution, but it makes it extremely difficult to extract the monochromatic lenslet PSF.

In order to measure the monochromatic lenslet PSFs, we have obtained calibration images in which the detector was uniformly illuminated by a single laser at a time.  The right panel of Fig.~\ref{fig:laser_flats} shows an inset of one such image.  The lenslet PSFs are now separated by $\approx$15 pixels, enabling their measurement out to $\approx$7 pixels from the center.  

We followed the same approach as \citetalias{Brandt2017} in order to extract the lenslet PSF positions and to extract the oversampled PSFlets.  Briefly, we began with the coordinates of the lenslets.  
The side-length $t$ of the hexagons is 7.4 mas (see Table \ref{tab:ifs_properties}) and the hexagon centers can be represented in a rectilinear grid for lenslet number $(i, j)$ as
\begin{equation}
\label{eq:lenslet_coords}
    \begin{split}
        x_{ij} &= t \sqrt{3} \left(i + 0.5 \, (i \, {\rm mod}\,2) \right), \\ 
        y_{ij} &= t \, j.
    \end{split}
\end{equation}
This corresponds to the representation internal to the pipeline.
We then assumed a wavelength-dependent 2D third-order polynomial mapping between the $x_{ij}$ and $y_{ij}$ lenslet coordinates from Equation \eqref{eq:lenslet_coords} and the PSFlet locations on the detector. This differs from \citetalias{Brandt2017} only in the coordinates of the lenslets themselves, which in this case reflect the hexagonal geometry of the SPHERE lenslets (CHARIS has square lenslets).

We lightly smoothed the monochromatic background-subtracted laser images by convolving them with a narrow Gaussian and then fit for the polynomial coefficients that maximize the peak intensity at the fit position of each lenslet on the detector.  In other words, we determined the polynomial functions of the lenslet coordinates ($x_{ij}, y_{ij}$) that give the most accurate pixel locations of the laser spots shown in Fig.~\ref{fig:laser_flats}.  These polynomials then give the mapping between lenslet coordinate and PSFlet centroid on the detector at a given wavelength.  

After fixing the mapping between lenslet and PSFlet location, we used an approach similar to that of \citet{Anderson+King_2000} to derive an oversampled effective PSF.  Briefly, we constructed a grid that was oversampled by a factor of 9 and placed each empirical PSFlet at the appropriate position within this grid.  We then took this sparsely sampled grid, convolved with a Gaussian to smooth it, and then performed a deconvolution.  This step accounts for the varying signal-to-noise ratios (S/N), where some offsets are sampled more thoroughly than others depending on the exact positions of the PSFlets.  We constructed the empirical oversampled PSFlets separately in 25 regions of the detector to account for potential spatial variations of the PSFlets.  We refer to \citetalias{Brandt2017} for additional details.  

\begin{figure}
\includegraphics[width=\linewidth]{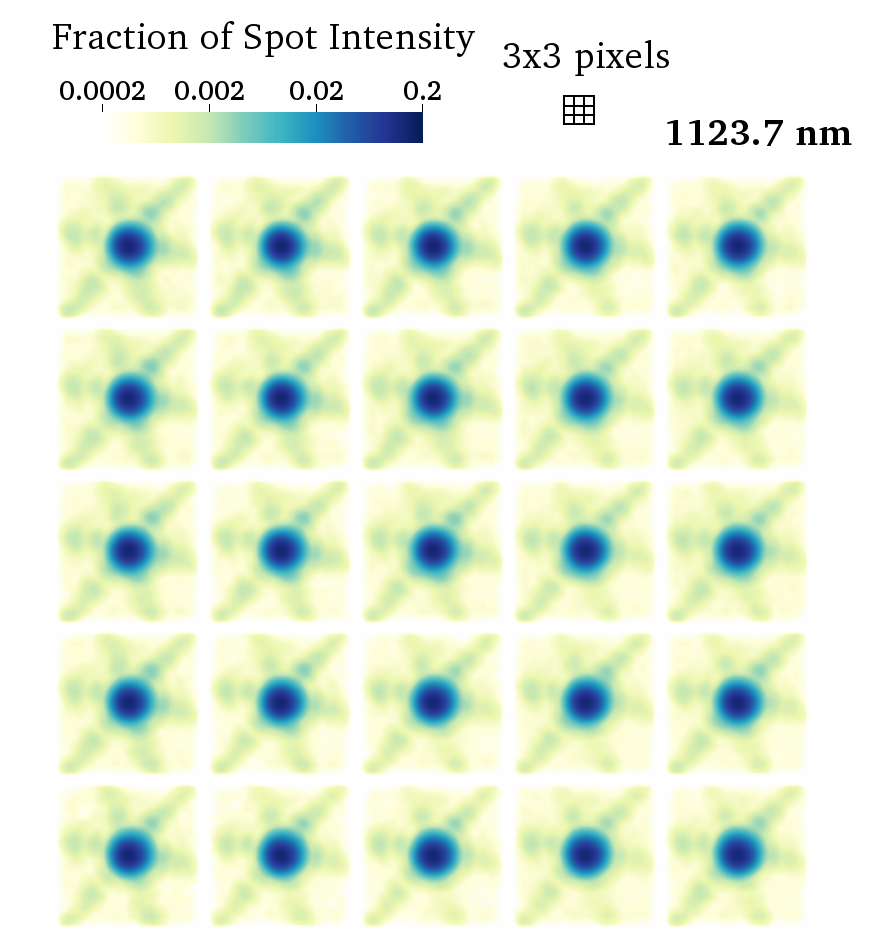}
\caption[Oversampled lenslet point-spread-functions]{Oversampled PSFlets at 1123.7 nm reconstructed from monochromatic laser images like that shown in Figure~\ref{fig:laser_flats} and normalized to unit intensity after pixel sampling. The detector is divided into 25 regions, each panel corresponding to the associated location on the detector: the top left PSFlet corresponds to the top left corner of the detector, etc. A $3\times 3$ pixel grid is shown as a size comparison. Similar libraries are created at the other three calibration wavelengths and included with the pipeline. The oversampled PSFlets can be used to construct a model of the pixellated microspectra corresponding to monochromatic or broadband light imaged by the lenslet array. The six diffraction spikes result from the hexagonal lenslet geometry, while the PSFlet shapes are extremely homogeneous across the FoV and give no indications of a changing focus.}
\label{fig:lenslet_psfs}
\end{figure}

Figure~\ref{fig:lenslet_psfs} shows our resulting oversampled lenslet PSFs.  The PSFs have a central core surrounded by six diffraction spikes; these result from the hexagonal lenslet geometry.  The PSFlets are more symmetric than those of CHARIS (\citetalias{Brandt2017}) and vary little across the FoV.  This suggests a uniform focus for the SPHERE lenslets throughout the FoV.  These lenslet PSFs include both optical and pixel sampling effects.  Interpolating them gives an empirical PSFlet at each position.  In practice, we use bilinear interpolation spatially to obtain the PSFlet for a given lenslet, and then interpolate within the oversampled PSFlet to obtain model monochromatic PSFlets.  These may be used to measure the instrumental profile for optimal extraction or to reconstruct the microspectra themselves for a least-squares extraction.

The instrument profile, that is, the PSFlet profile perpendicular to the dispersion direction, is required for optimal extraction.  The oversampled PSFlets make this straightforward to determine.  The instrumental profile is simply the PSFlet collapsed along the dispersion direction (vertical in Figure \ref{fig:lenslet_psfs}).

\subsection{Wavelength calibration} \label{sec:wavecal}

\begin{figure}
\centering
\includegraphics[width=\columnwidth]{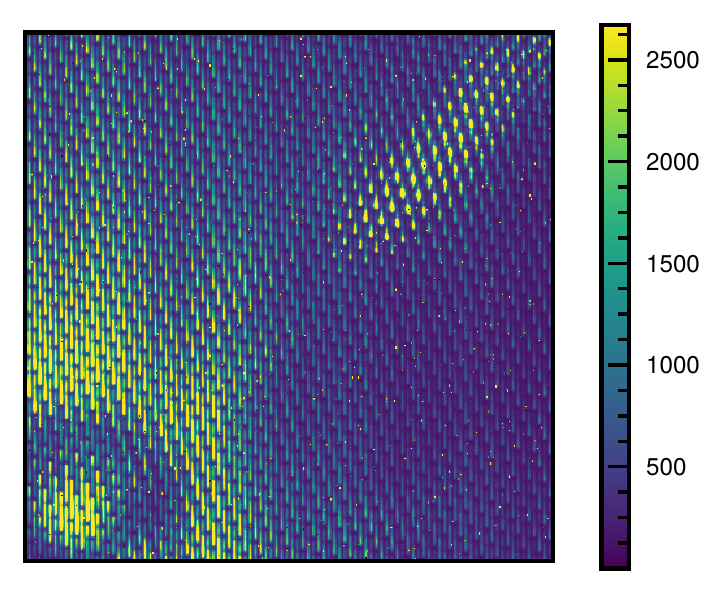} 
\caption[Cutout of unprocessed IFS raw data of 51~Eri~b]{Cutout of unprocessed IFS raw data of 51~Eri~b in YH-mode with satellite spots induced by the deformable mirror. The dispersion is in vertical direction from longer wavelengths at the top to shorter wavelengths at the bottom. The star center is in the bottom left corner.}
\label{fig:ifs_raw_data}
\end{figure}

Figure~\ref{fig:ifs_raw_data} shows an inset of a raw SPHERE IFS image; the image consists of microspectra arranged in the geometry of the lenslet array. One of the most important steps in the data reduction is therefore assigning each pixel membership to a specific spectrum (lenslet) and the wavelength corresponding to the position in the dispersed spectrum.  We refer to these steps collectively as obtaining the wavelength solution.  

Section~\ref{subsec:psflets} describes the extraction of oversampled PSFlets from a series of monochromatic laser images.  The first step of extracting these PSFlets is to derive the mapping between lenslet and detector position at a given wavelength.  This results in anchors to the wavelength solution at either three (in YJ-mode) or four (in YH-mode) wavelengths.  We treat the full wavelength solution as a quadratic in $\log \lambda$ for YJ-mode and a cubic in YH-mode.  This results in interpolation in both cases.

\begin{figure}
    \centering
    \includegraphics[width=\linewidth]{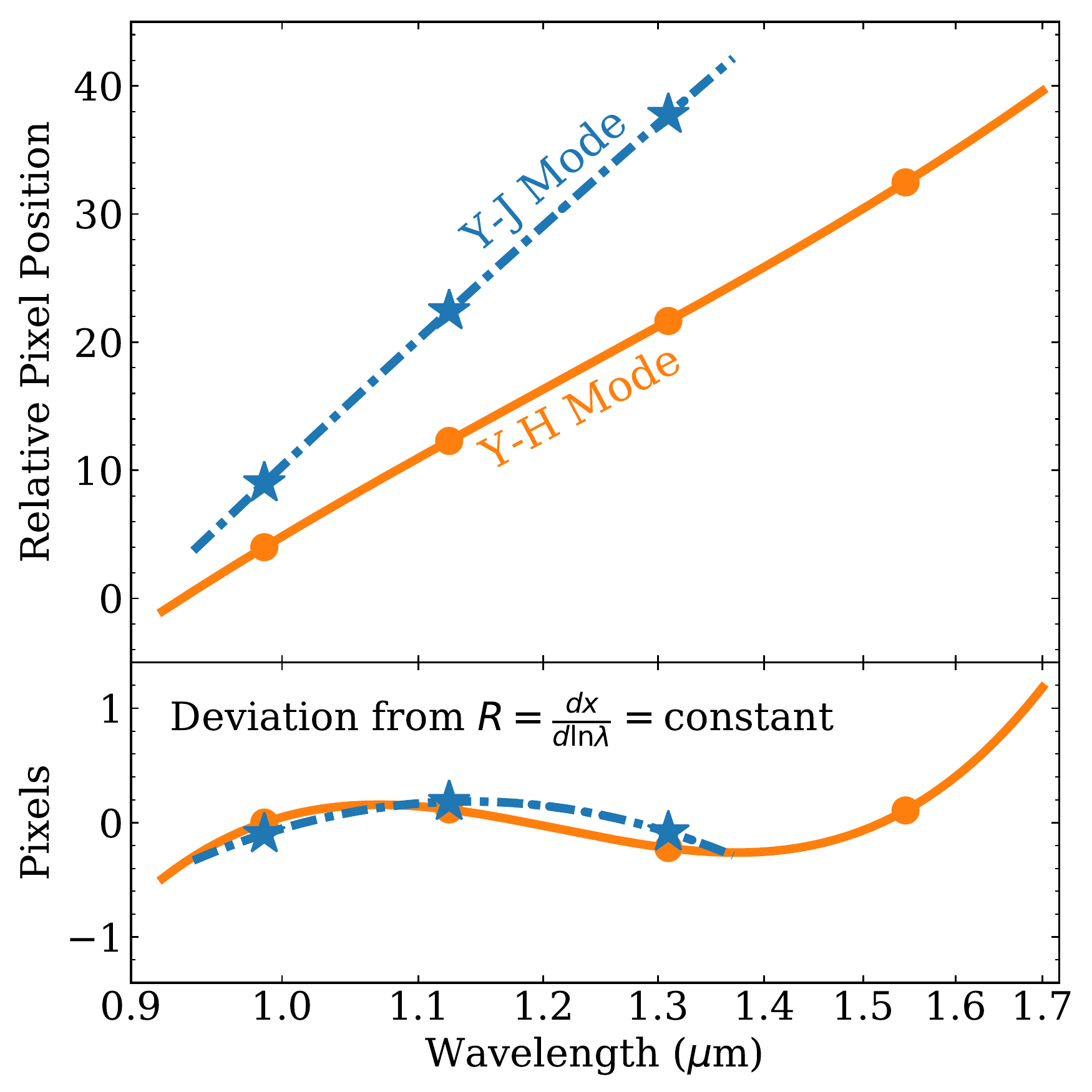}
    \caption{Wavelength solution derived from the positions of the three (in YJ-mode) or four (in YH-mode) calibration lasers for the central lenslet.  The YJ and YH lines are offset from one another for clarity.  Bottom panel: Deviations from a constant spectral resolution, i.e., a constant number of pixels per logarithmic wavelength (the straight line in the top panel).}
    \label{fig:wavesol}
\end{figure}

Figure~\ref{fig:wavesol} shows the measured PSFlet positions in the center of the detector at the laser calibration wavelengths for the YJ- and YH-modes.  It also shows the deviations from constant spectral resolution, that is, a constant number of pixels per logarithmic unit of wavelength.  These deviations reach a maximum of $\approx$0.3 pixels, or about 0.005\,$\mu$m, across most of the instrument spectral range.  They reach $\approx$1 pixel, or about 0.02\,$\mu$m, at the long-wavelength extreme of the YH-mode bandpass.  The lack of a calibration laser at a wavelength longer than 1.54\,$\mu$m means that the wavelength solution in this spectral range is an extrapolation and should be treated as uncertain.

The actual wavelength solution used by the pipeline includes the full third-order 2D polynomial fit described in Section~\ref{subsec:psflets}.  This gives a PSFlet position for each lenslet and each wavelength.  Each lenslet then has either three (in YJ-mode) or four (in YH-mode) pixel-wavelength ($x,\lambda$) pairs.  The unique second- or third-order polynomial $x(\ln \lambda)$ then is the wavelength solution.  This part of the wavelength solution is distributed with the pipeline and treated as static: it is unchanged for all reductions of SPHERE IFS data.
The first step with a new set of raw SPHERE data is to determine the spatial shift from the static wavelength solution described above.  We assume that this shift is small and can be represented by a horizontal shift, a vertical shift, and a rotation angle.  For this step, we use the images taken as part of the standard SPHERE observing sequence, with either three or four calibration lasers.  We first mask bad pixels in these images and then lightly smooth them with a Gaussian.  We then obtain the pixel values at the locations tabulated in the static wavelength solution for all lenslets.  The sum of the smoothed images at all of these locations is our figure of merit.  We apply a shift in $x$, a shift in $y$, and a rotation to these positions in order to maximize this figure of merit.  The best-fit shifts are small in practice: $\lesssim$1\,pixel and $\lesssim$10$^{-4}$\,radians. These shifts are then combined with the static calibration data to describe the full wavelength solution for a given set of data. Small shifts are correlated with the temperature of the instrument due to thermal expansion and contraction. Larger shifts can occur rarely after an earthquake occurs close to Paranal. 

\subsection{Calibrating a raw frame}

The preceding sections described the process of constructing a bad-pixel mask, a detector flat-field, a background template, a wavelength solution, and oversampled PSFlet models.  All of these are static and distributed with the pipeline, except for a small shift in the wavelength solution for each set of observations.  For an individual raw SPHERE IFS image, we use some of these data products before proceeding to spectral extraction.  In this section, we describe these initial steps.

Our first step in reducing a raw SPHERE IFS frame is to mask bad pixels.  We then subtract a background using our background template: we use least-squares fitting to the four regions in each corner of the detector that the lenslets do not illuminate (as described in Section~\ref{sec:background_template}).  This produces a background-subtracted masked image.

We do not perform flat-fielding at this stage, but defer that step to spectral extraction.  With a background-subtracted image, masked bad pixels, and a wavelength solution appropriate to a given frame, we proceed to the spectral extraction step.  We defer the discussion of the lenslet flat-field to Section~\ref{subsec:lenslet_flat} because constructing this flat-field requires spectral extraction of white-light calibration images.

\section{Extracting a data cube}
\label{sec:extraction}
In this section we present a step-by-step discussion of the process of extracting a spectral data cube from raw data using our adaptation of the CHARIS pipeline.
A SPHERE IFS frame, after background removal, consists of $\approx$20,000 microspectra that are arranged in a grid across the detector.  The core step in the data reduction is to extract each of these spectra and reconstruct the 3D $(x,y,\lambda)$ data cube.
The wavelength solution, described in Section~\ref{sec:wavecal}, gives the location, or trace, of each microspectrum.  Each is $\approx$39 pixels long on the detector, and the dispersion direction is very nearly parallel to the vertical dimension of the detector.  The trace of each spectrum is separated by $\approx$5 pixels from its nearest neighbor.
 
The SPHERE IFS pipeline described here implements three approaches to extract the microspectra: boxcar extraction, optimal extraction, and least-squares extraction.  The DRH (ESO) pipeline, as well as the GPI pipeline \citep{Perrin2014}, implement only boxcar extraction.  In this section we summarize each approach in turn. The approaches and algorithms we use are substantially identical to those use for CHARIS; we refer to \citetalias{Brandt2017} for detailed discussions.  

\subsection{Boxcar extraction}

The first spectral extraction technique implemented for SPHERE is boxcar extraction, a technique similar to aperture photometry.  In this approach, the pipeline adds all flux perpendicular to the dispersion direction out to a certain number of pixels from the trace.  With a spacing of $\approx$5 pixels between traces, a typical choice would be to add the flux in a 5-pixel-wide aperture centered on a given spectrum's trace (2 pixels out in each direction).  The total flux is then assigned to the wavelength of the central pixel in this box, the one along the trace.  This wavelength is known from the wavelength solution.  Each microspectrum is then defined on a different wavelength array: the wavelength corresponding to integer pixels along the trace of the microspectrum.  The final step is then to interpolate these microspectra onto a common wavelength array.  

Boxcar extraction has several important limitations.  The first is that it has no way of treating bad pixels.  Bad pixels must be corrected for and then treated exactly as if they were good pixels.  The second limitation is that it applies the same weight to flux in the core of the trace, where the S/N is higher, as to the outskirts, where it is lower.  A wider aperture captures more of the flux, but at the expense of S/N.  This limitation is minor in the limit of a high S/N, but when read noise and background noise are significant, boxcar extraction entails a significant S/N penalty relative to other approaches.  Finally, boxcar extraction lacks a straightforward way of treating crosstalk, where light from one microspectrum contributes flux to the aperture around a neighboring microspectrum.  The DC and \texttt{vlt-sphere} pipelines implement a crosstalk correction before boxcar extraction.  

\subsection{Optimal extraction}
\label{sec:optimal_extraction}
Other spectral extraction approaches offer a superior performance to boxcar extraction if the spectral profile perpendicular to the trace is known.  In Section~\ref{subsec:psflets} we describe the calculation of this profile using monochromatic PSFlets out to $\approx$7 pixels from the trace.  This captures the profile $\approx$2 pixels past the trace of the neighboring spectrum.  This provides a way to implement optimal extraction \citep{Horne1986} and least-squares extraction \citep{Brandt2017}, and to implement a crosstalk correction by iterative modeling of the full 2D array of microspectra.

The implementation of an optimal extraction in the SPHERE IFS pipeline works similarly to boxcar extraction, but weights the flux from each pixel by the profile calculated from the \mbox{PSFlets}.  We include the pixel flat-field in this profile for each lenslet, that is, the pixel response is included in the model rather than in the data.  

Each spectrum contains light from its neighbors, both along and perpendicular to the dispersion direction. This light is often called spectral crosstalk. We remove crosstalk in optimal extraction by first performing a least-squares extraction as described in the following subsection.  This involves a full modeling of the 2D array of microspectra, including an attribution of light to each lenslet.  We can then remove the light extending more than 2 pixels from the trace of each microspectrum before performing optimal extraction.

After optimal extraction, each microspectrum remains defined on its own unique wavelength array corresponding to the central wavelengths of each pixel along the trace.  In the same way as for the boxcar extraction, we take the final step of interpolating all microspectra onto a common wavelength array.

\subsection{Least-squares extraction}
\label{sec:least_squares_extraction}

The final extraction technique implemented in our SPHERE IFS pipeline is least-squares extraction, which was successfully implemented for CHARIS by \citet{Brandt2017}.  In this approach, we model each microspectrum in 2D.  We take the model spectrum incident on a lenslet to be locally flat over short ranges of wavelength. The actual wavelength range corresponds to one spectral resolution element; we discuss the choice of extracted spectral resolution below.  We then assume that the full spectrum is a linear combination of these top-hat spectra in wavelength space.  Because we know the lenslet PSFs, these top-hat spectra over small ranges of wavelength correspond to PSFlets in detector space that have been dispersed by one spectral resolution element.  A set of coefficients for the amplitudes of these slightly dispersed PSFlets corresponds to a model microspectrum on the detector. Extracting these coefficients becomes a least-squares problem of fitting the microspectra themselves.  Least-squares extraction implicitly deconvolves the spectra with the line-spread function.  Because the top-hat spectra are defined over the same wavelength ranges for each microspectrum, least-squares extraction does not require interpolation onto a common wavelength array.  

We adopted the same procedure for crosstalk correction in least-squares extraction as for optimal extraction.  This procedure was also described in \citet{Brandt2017}. We first extract each microspectrum using the procedure described in the previous paragraph.  We model each microspectrum using the dispersed PSFlets, fitting the model microspectra over a region extending 2 pixels perpendicular to the trace in each direction. These microspectra include light from their neighbors as spectral crosstalk.  Our model PSFlets, however, extend 7 pixels from the trace.  We compute the 2D image expected from our extracted spectra, now adding the light out to this larger extent.  
This full image has more photons than the actual measured image because photons from several lenslets contributed to the flux in the core of each microspectrum.  For optimal extraction, we subtract the observed detector image from the full model image; this gives the overcounting of photons due to spectral crosstalk.  We then subtract this residual from the actual image before performing optimal extraction.  For the least-squares extraction, we fit this residual 2D array of microspectra exactly as we did for the original array of microspectra. We then subtract the residual 1D spectra from the previously extracted 1D spectra to perform the crosstalk correction.

Least-squares extraction has a number of advantages over other approaches in principle.  It models the full 2D detector image, allowing it to account for (and fit) read noise patterns (\citetalias{Brandt2017}).  Least-squares extraction can naturally account for bad or missing data, using the true uncertainties on the measured flux of each pixel.  It also performs a deconvolution of the microspectra with the line-spread function.  This deconvolution comes at a price, however.  The least-squares extracted microspectra are covariant at neighboring wavelengths, and this covariance can become very strongly negative when attempting to deconvolve to a higher spectral resolution.  The SPHERE IFS pipeline extracts the microspectra at a resolution $R = 55$ for YJ-mode and $R=35$ for YH-mode, corresponding to $\approx$2 pixels along the dispersion direction.  This represents a compromise: A higher spectral resolution will enable better modeling of the detector image and return a higher-resolution data cube, but at the cost of stronger covariances between neighboring wavelengths.  To save disk space, the pipeline does not save the full covariance matrix for each spectrum (which would multiply the file size by more than a factor of 10).  Depending on the needs of the user and on the properties of a data set, either optimal extraction or least-squares extraction is likely to offer the best-quality data cubes.

\begin{figure*}
     \centering
     \begin{subfigure}[b]{0.48\textwidth}
         \centering
         \includegraphics[width=\textwidth, clip]{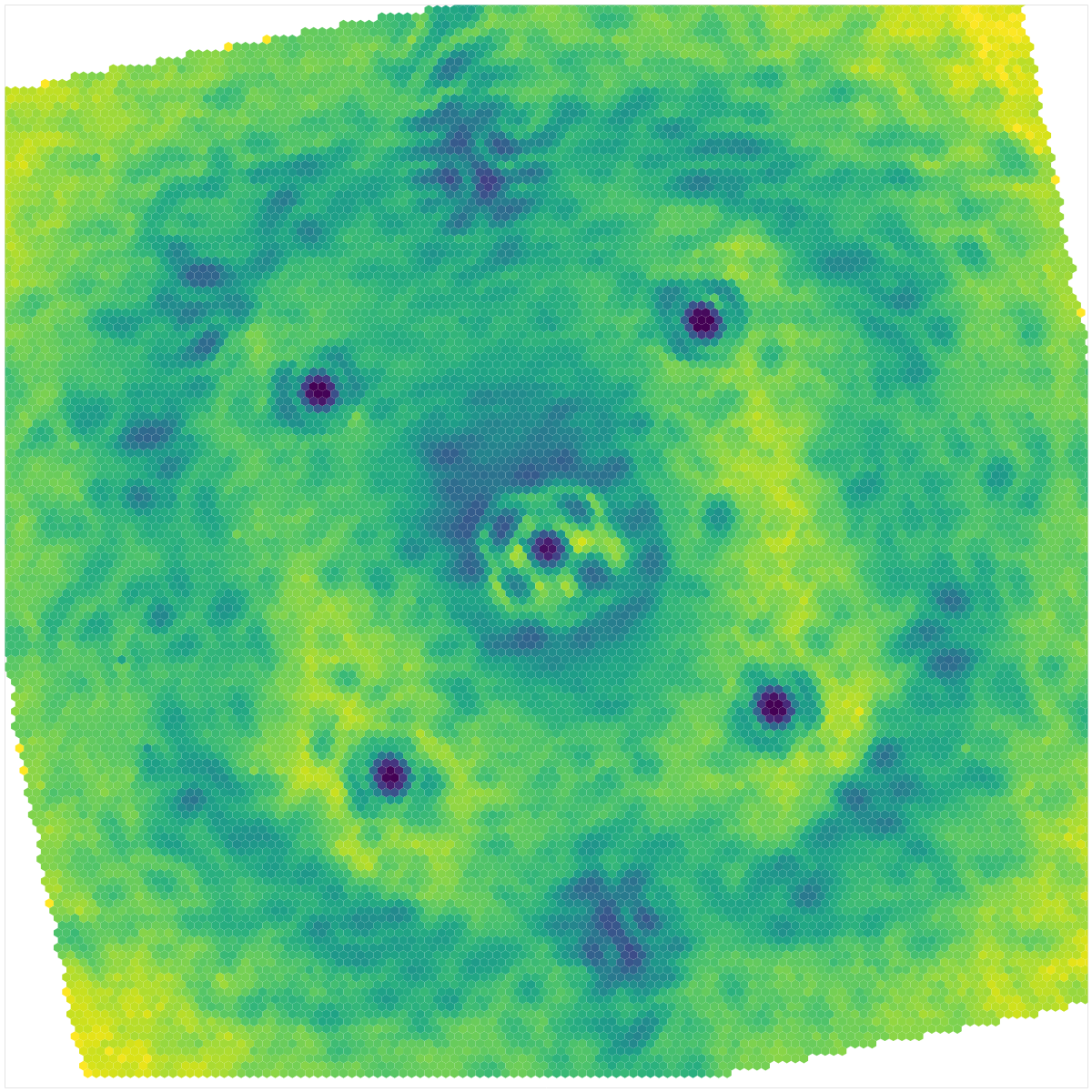}
         \caption{Least squares}
         \label{fig:least_square_image}
     \end{subfigure}
     \hfill
     \begin{subfigure}[b]{0.48\textwidth}
         \centering
         \includegraphics[width=\textwidth, clip]{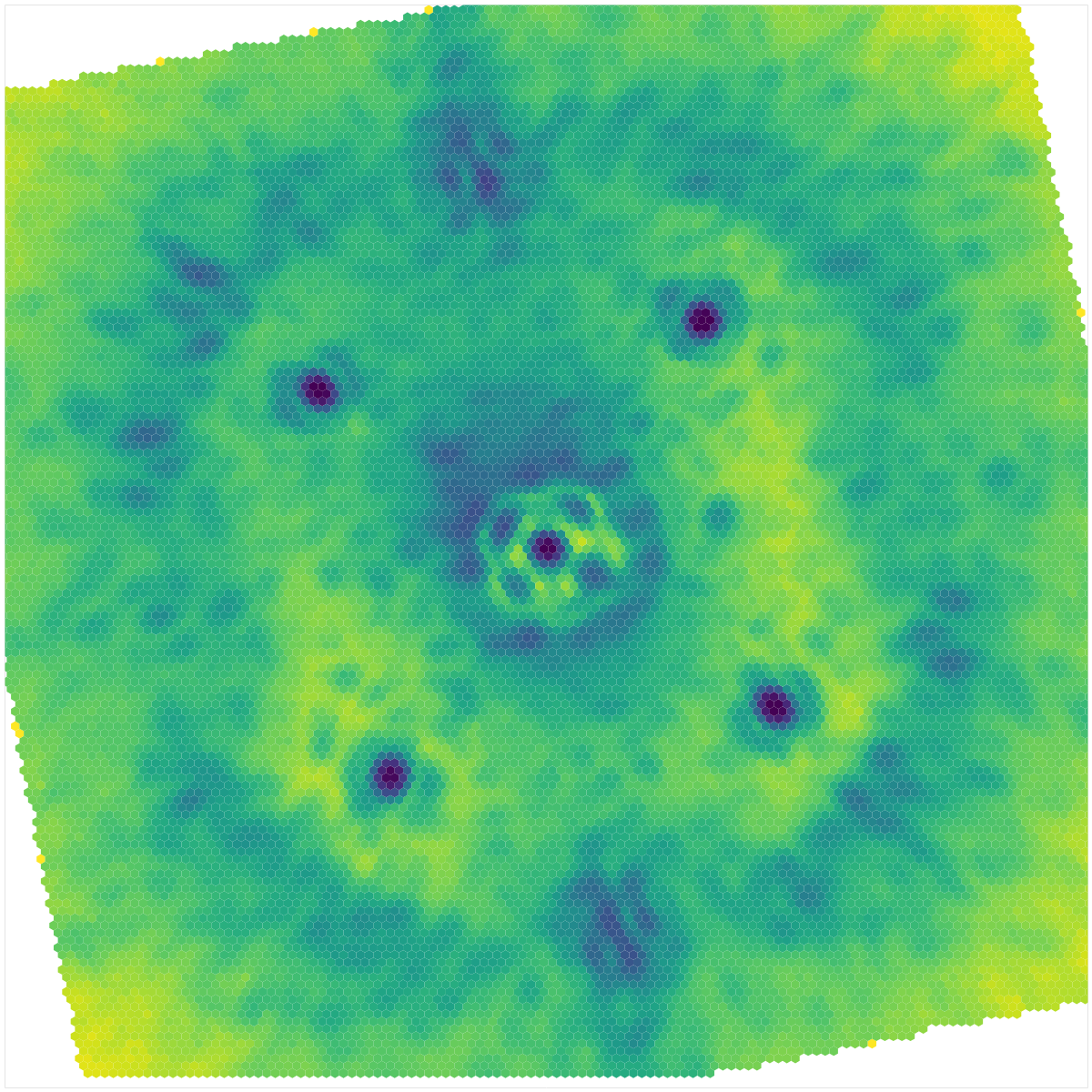}
         \caption{Optimal extraction}
         \label{fig:optimal_extraction_image}
     \end{subfigure}
        \caption{Single slice through the same YH-mode data cube extracted two ways and shown in the original hexagonal geometry: (a) Least-squares extraction, and (b) optimal extraction. Least-squares extraction performs a deconvolution with the line-spread function and as such appears less smoothed in general, but may appear more noisy in the low-S/N regime for this reason.}
        \label{fig:extraction_methods}
\end{figure*}

Figure~\ref{fig:extraction_methods} shows an extracted image using the two main methods. The speckle pattern using the least-squares method is less blurry because it performs a deconvolution with the line-spread function and does not require interpolation in the spectral dimension. Both images are shown in the original hexagonal geometry in which the data are taken.

\subsection{Lenslet flat-field}
\label{subsec:lenslet_flat}

The lenslet flat-field is distinct from the detector flat-field we described in Section \ref{sec:detectorflat}.  The lenslet, or IFU, flat-field tells us the relative throughput of each lenslet.  The SPHERE DRH pipeline uses images in which the lenslet array is uniformly illuminated by white light to construct this flat-field.

The CHARIS instrument uses the wavelength-averaged (near-)monochromatic spot amplitude patterns obtained with the supercontinuum source and a tunable narrow-band filter directly as an indicator for the lenslet transmission. Unfortunately, this is not possible for SPHERE as the lasers are coherent light sources and susceptible to fringing and spurious amplitude variations across the lenslet array. This is not an issue when we make the PSFlet library as the models are normalized, but they are not reliable indicators for absolute transmission. We therefore used the same approach as the SPHERE DRH pipeline ``instrument flat-field'' step, which corresponds to our lenslet flat-field, and extracted the spectra for a white lamp illuminating the IFS using the optimal extraction algorithm described above.  We then normalized by the average of all lenslets and divided the output data cube by the wavelength-averaged resulting flat-field.

\subsection{Lenslet geometry}

The native lenslet geometry of the SPHERE IFS is hexagonal.  This differs from the similar CHARIS and GPI IFSs, which both contain square lenslet arrays.  It also differs from the rectilinear geometry that is most natural to a numerical array.  For this reason and for compatibility with the many software tools now available to postprocess high-contrast imaging data \citep[e.g.][]{pyklip2015, Cantalloube2015, VIP2016, specal2018, Dahlqvist2020, Samland2021}, the SPHERE DRH interpolates data cubes spatially onto a rectilinear grid.

The SPHERE IFS pipeline that we present here offers two options.  The first is to interpolate the data cubes onto the same rectilinear grid geometry as that used by the SPHERE IFS pipeline.  The second option is to preserve the native hexagonal lenslet geometry, which means that each physical hexagonal lenslet corresponds to an individual hexagonal spaxel in the output image. In this image, the midpoint of each spaxel corresponds to a position on a hexagonal grid. The side-length $t$ of the hexagons is 7.4 mas (see Table 1), and the orientation of the hexagon grid is in the so-called pointy topped configuration, as opposed to flat topped. The hexagon centers are given by Equation~\eqref{eq:lenslet_coords}.

The hexagonal grid is resampled onto a rectilinear grid using the Sutherland-Hodgeman algorithm, which can be used to compute the area of overlap between two polygons (in this case, the original hexagon and each pixel of the new rectilinear grid). The area of overlap between pixels of the two grids are only computed once and saved in a calibration file that is provided with the pipeline. The areas of overlap are then used to divide the flux of the original image onto the rectilinear pixel grid.

We include routines that allow a user to visualize the extracted cubes in the native hexagonal geometry without resampling, as most normal image viewers like \textsc{ds9} can only view images with square pixels. A command-line script allows quick access to viewing an extracted data cube, similar to the other routines of the package. It allows the interactive viewing of an extracted image cube with a slider for wavelength.

\subsection{Astrometric and spectrophotometric calibration}

After the spectral cube is extracted, we extract the location of the occulted central star using the four satellite spots, which are generated using the deformable mirror for this purpose. We fit a 2D Gaussian to each spot in the resampled images and compute the point at which the lines connecting the spots on opposite side of the star cross to obtain the image center for each wavelength. For this purpose, we adapted the routines found in \citet{Vigan2020_pipeline}. The image center positions are then fit by a polynomial of suitably high order in the wavelength dimension (second order in x-direction and third order in y-direction) to obtain the final result \footnote{More information about the astrometry and optical distortions of the instrument can be found in \citet{Maire2016b}.}. The first and last wavelength channel and the channels that are most strongly affected by atmospheric transmission ($\approx$1.14 and $\approx$1.37 micron, see, e.g., the lower panels of Fig.~\ref{fig:wd_spectrum}) are excluded from the fit. These channels have a lower S/N due to low atmospheric transmission, and in the case of the first and last channels, low instrumental transmission. The measured position can be biased by the gradient in the transmission profile, which is significant in all of these channels. These channels, due to their inherently lower S/N, are also more prone to residual spectral crosstalk. The crosstalk modeling is discussed in Section~\ref{sec:least_squares_extraction}.

For the spectrophotometric calibration, we used the unsaturated noncoronagraphic PSF images obtained before and after the coronagraphic sequences. We accounted for the neutral density filter throughput at each wavelength and adjusted the integration time to match the coronagraphic observations. We performed aperture photometry on the PSF ($r=3$ pix) and subtracted the sky background as determined by an annulus between $r=$15--18 pix. We then scaled the flux of all PSF images to match the two PSFs that are closest in time to the coronagraphic sequence. The average of the scaled and background-subtracted PSF is used as our planet model PSF in postprocessing.
We performed the same aperture photometry on the satellite spots to determine their flux and calibrate them by matching the counts of the first and last satellite spot image to the flux image closest in time, respectively. For observations that continuously use the satellite spots during the coronagraphic sequence, we have a fully flux-calibrated sequence that allows adapting the planet model for postprocessing for each frame and wavelength. For those sequences that only obtain satellite spots at the beginning and end of the coronagraphic sequence, we can still use this method to effectively move the flux calibration closer in time to the data that are used in postprocessing.

\subsection{Summary of changes to the original CHARIS pipeline}

The adaptation of the CHARIS pipeline \citetalias{Brandt2017} to the SPHERE IFS required a number of changes that we summarize here. The pipeline can now extract data cubes from either instrument.

The original CHARIS pipeline was written in Python 2.7; we have modernized the entire code-base to be compatible with new Python versions. To enable it to handle both SPHERE IFS and CHARIS data, we isolated all instrument-specific and hard-coded variables and moved them to a new {\tt Instrument} module. Depending on the data provided to the pipeline, the corresponding instrument class is instantiated and changes the behavior of the pipeline according to the instrument and instrument mode of the data. The pipeline behavior then reflects the lenslet geometry, wavelength range and sampling, instrument-specific header keywords, and properties of the observatory location and average atmospheric transmission. 

The construction of calibration files using the {\tt buildcal} script is substantially identical to the original CHARIS pipeline. The appropriate calibration templates are specified by an instance of the respective instrument class.  These calibration templates, including oversampled PSFlets, background templates, and both detector and lenslet flat-fields, were derived as described in Section~\ref{sec:calibration}.  The spectral extraction step incorporates the new undispersed background-fitting routine, which currently only works for SPHERE IFS. 
To account for the hexagonal lenslet geometry, we modified the bad-lenset detection routine, which cosmetically replaces outlier lenslets by its neighbors in the extracted spectral data cube, as there are six equally distant neighbors as opposed to nine unequally distant neighbors in the rectilinear geometry. Last, we implemented a routine to resample the hexagonal spaxels into a rectilinear grid and included tools to plot the spectral data cube in its native hexagonal geometry.

\section{Results}
\label{sec:results}

In order to show the performance of our adaptation of the CHARIS pipeline for SPHERE IFS data, we extracted image cubes and compared the results with the extracted image cubes obtained from the \texttt{SPHERE Data Center} reduction, which is based on the ESO DRH pipeline and additional refinement tools \citep{SPHEREDC}. We first briefly discuss the visual differences in the extracted images of the two pipelines and the two main extraction methods in our pipeline (optimal extraction and least-squares extraction).

We then proceed to show the performance of the pipeline by performing two tests. First, we use both pipelines to extract a high-S/N spectrum of the known white dwarf companion \object{HD 2133B} to a main-sequence star to ascertain the reliability of the spectral extraction requiring minimum postprocessing. Second, we perform the same ADI postprocessing on a high-contrast imaging sequence of the exoplanet \object{51 Eridani b} using both pipelines. The postprocessing parameters and pipeline \citep[TRAP,][]{Samland2021} are kept the same in both cases. We then compare the spectra and contrast curves.

\subsection{Visual comparison of extracted images}
\label{sec:visual_comparison}

\begin{figure*}
\centering
\begin{tabular}{ccc}
\includegraphics[width=6cm]{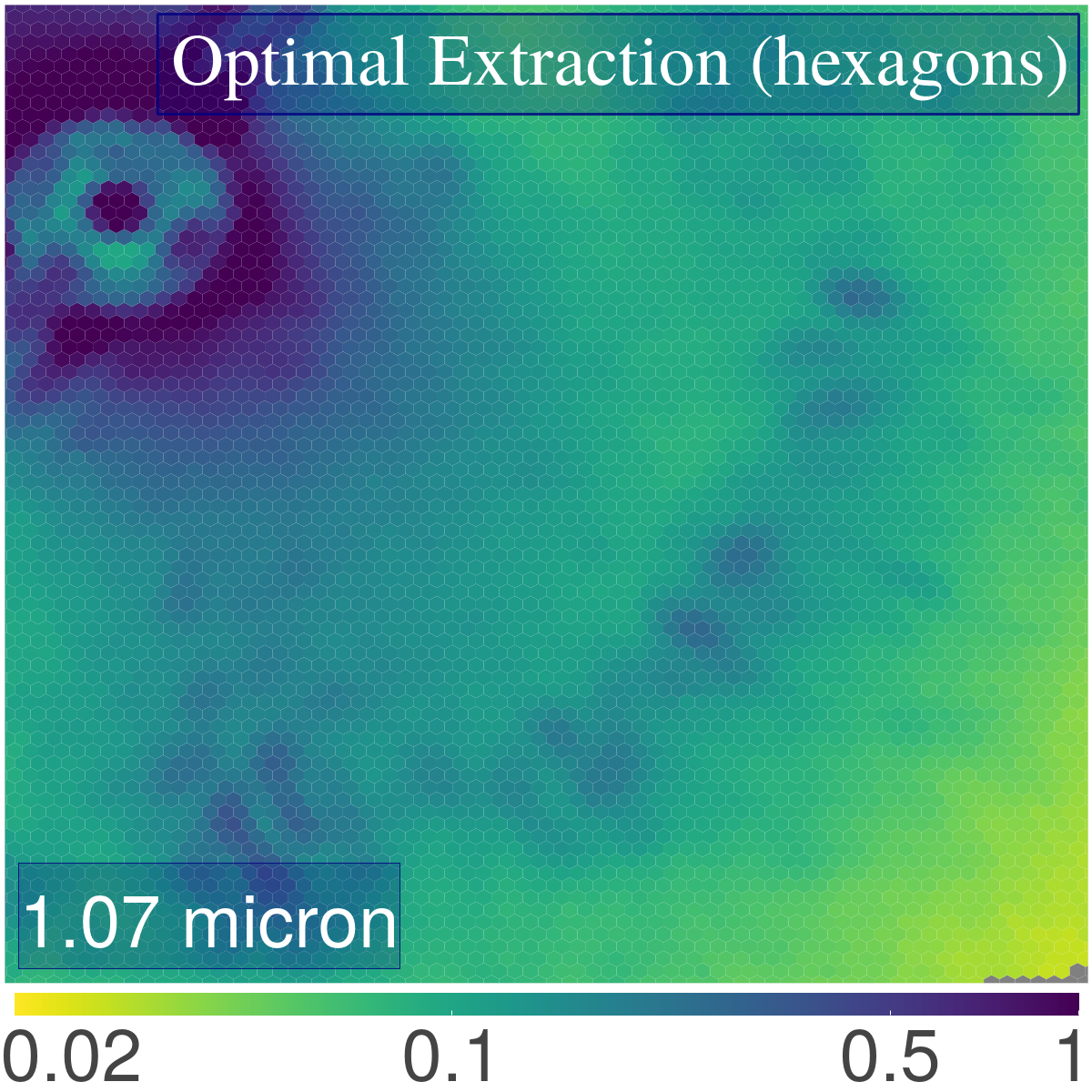} &
\includegraphics[width=6cm]{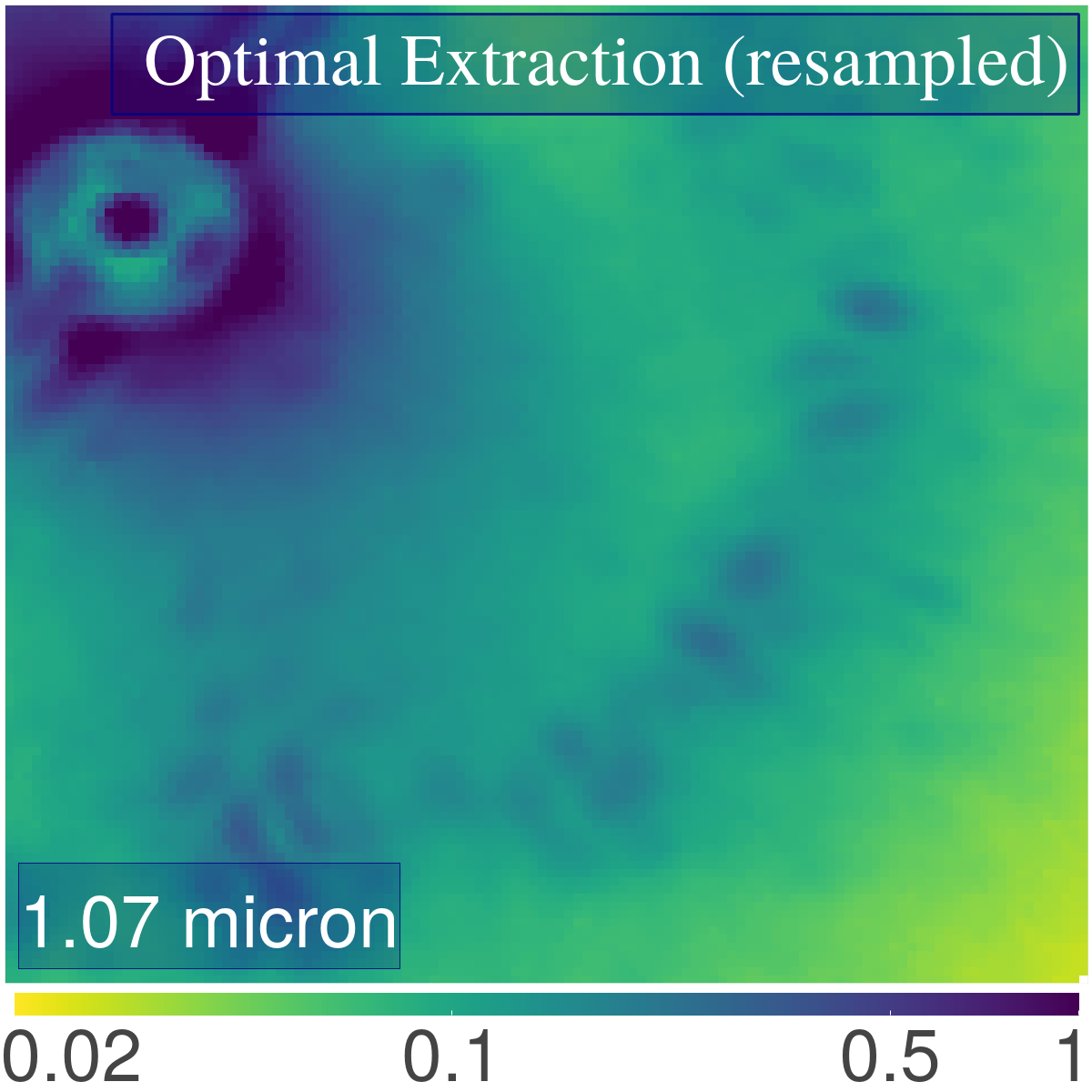} &
\includegraphics[width=6cm]{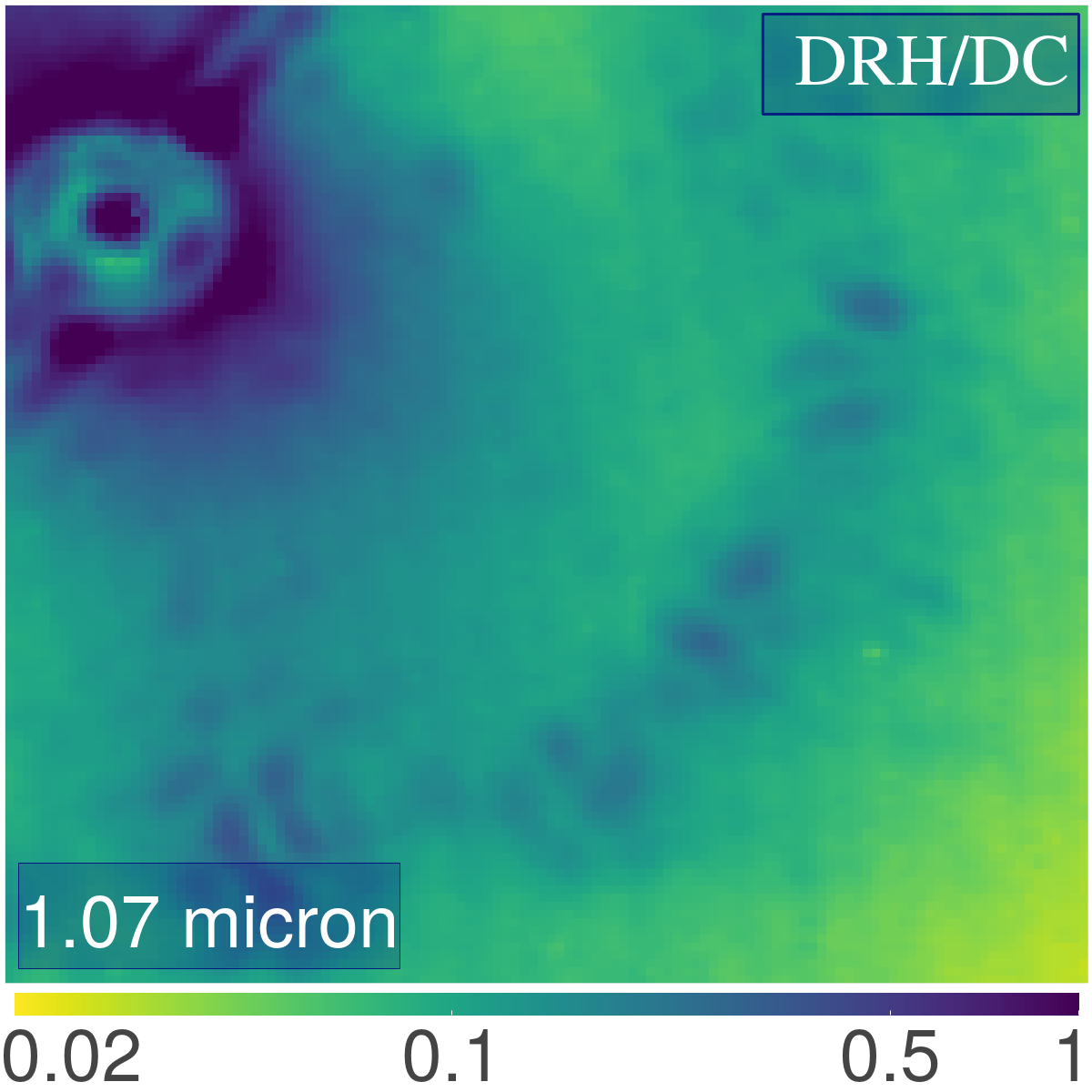} \\
\includegraphics[width=6cm]{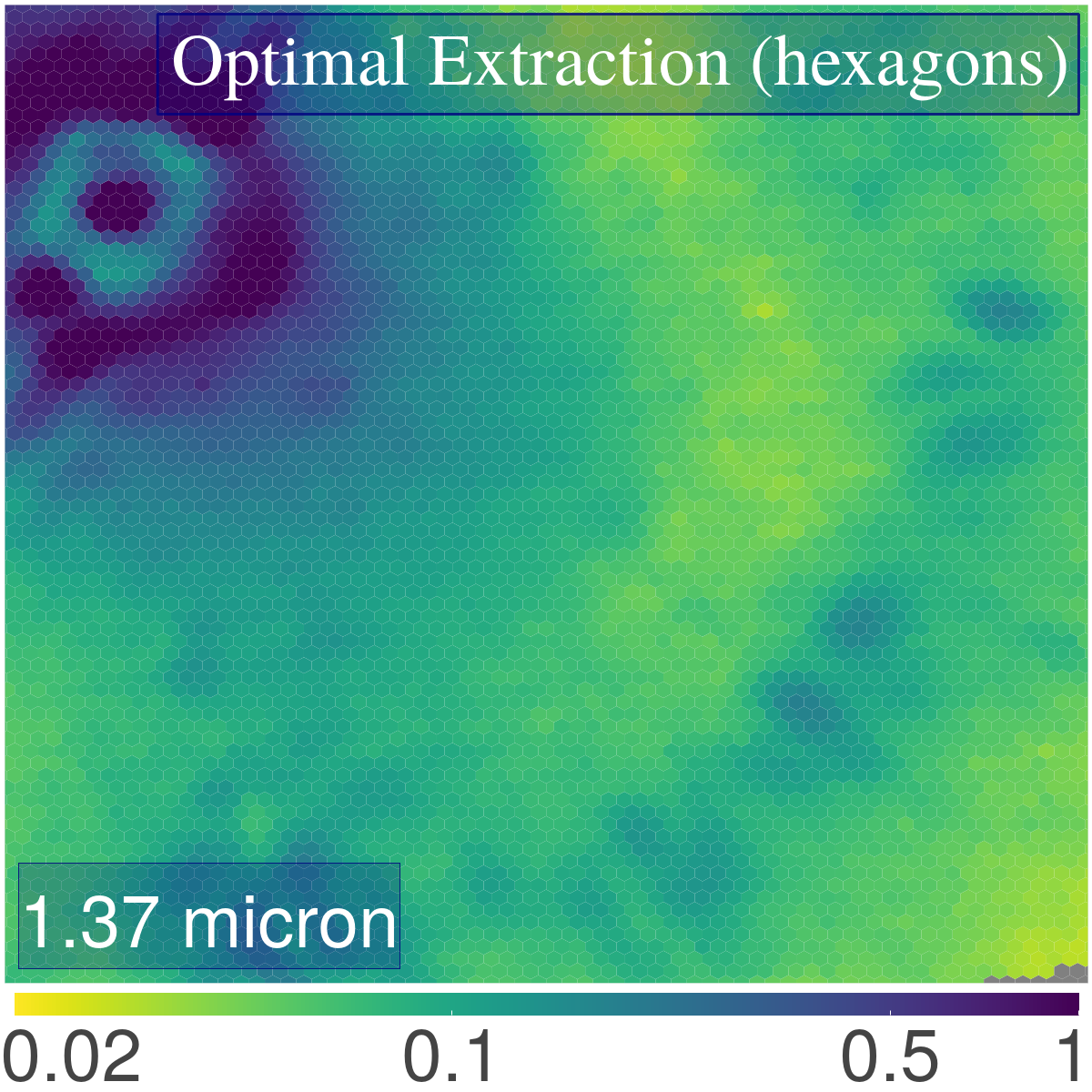} &
\includegraphics[width=6cm]{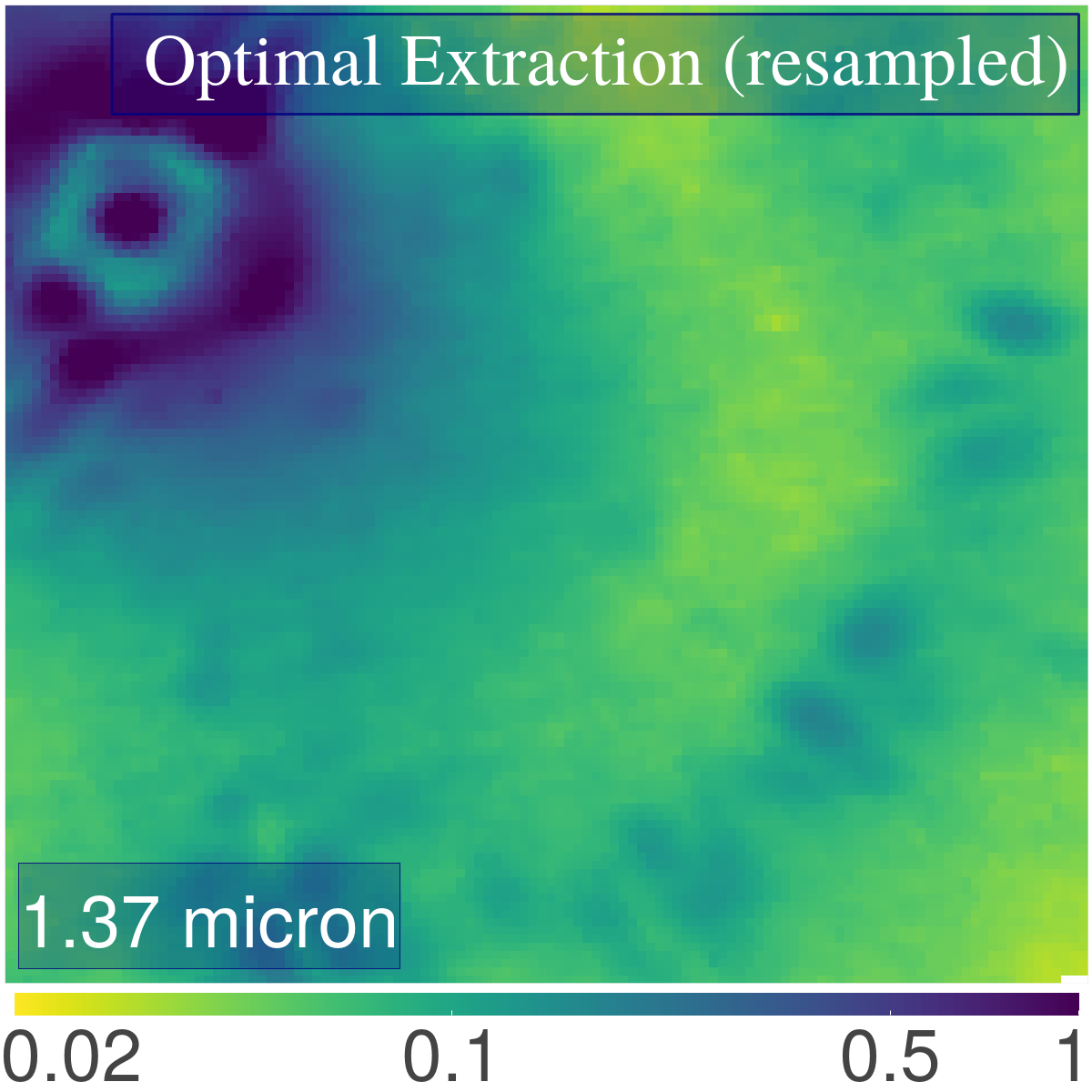} &
\includegraphics[width=6cm]{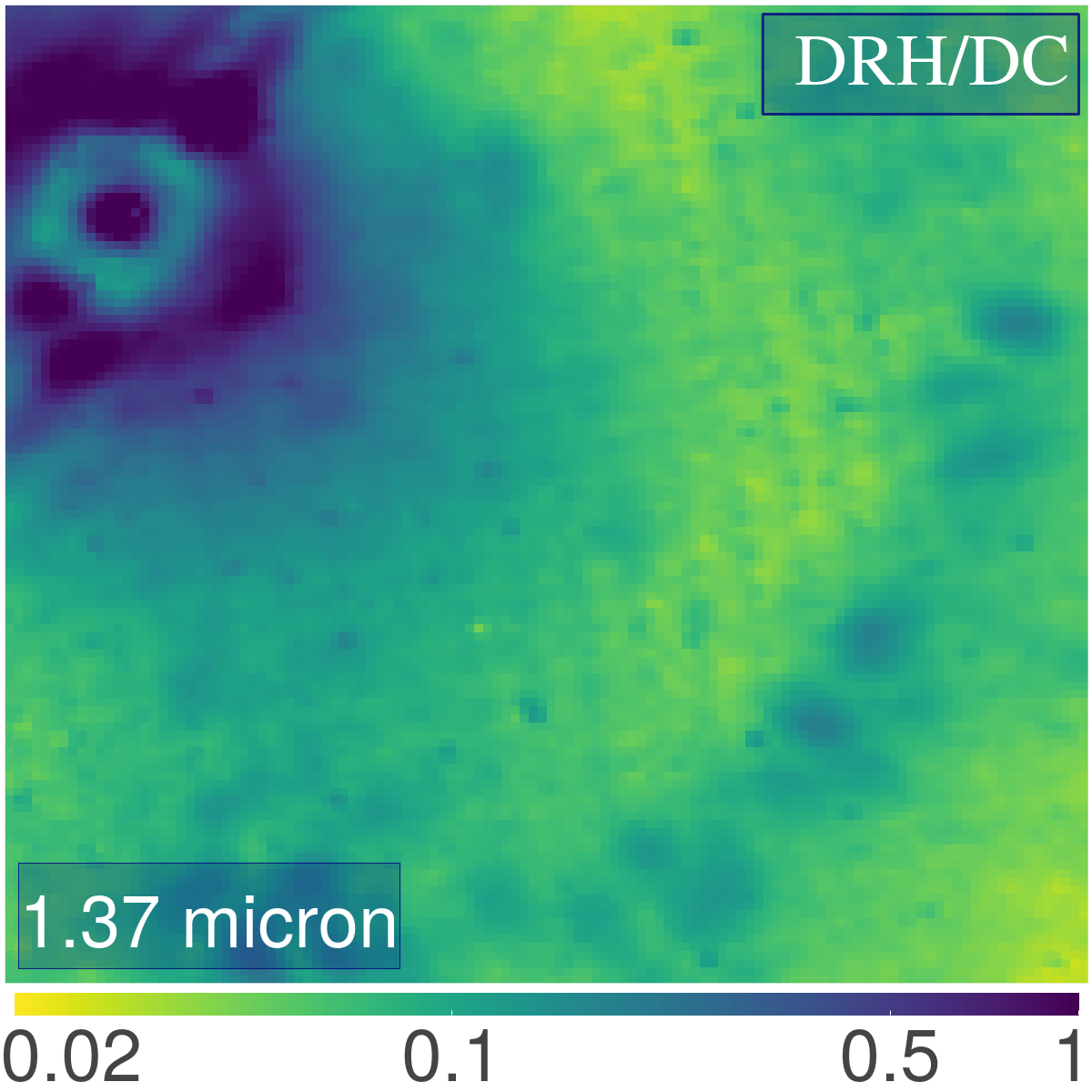}
\end{tabular}
\caption{Comparison of normalized image slices extracted from the 2015 51~Eri data set with our pipeline and the DRH/DC for two wavelengths: one at high S/N (top row) and one at low S/N (bottom row). Top row: At the peak of emission at around 1065~nm, shown in logarithmic scale. Bottom row: The same for the channel with the lowest S/N, at around 1370~nm.
From left to right: Optimal extraction frame (YH-mode) in hexagonal geometry; the same frame resampled to rectilinear grid; DRH pipeline reduction with additional routines for crosstalk correction and wavelength correction from the SPHERE DC.}
\label{fig:comparison1}
\end{figure*}

As a first step, we extracted a single wavelength slice from a data cube in the 51~Eri~b SPHERE IFS data set taken on the \mbox{2015 September 25} \citep{Samland2017} in the YH-mode.
The data set was taken in below median conditions and shows a relatively strong low-wind effect \citep{Couder1949, Sauvage2016, Milli2018SPIE_Wind}. Figure~\ref{fig:comparison1} shows two wavelength slices (top: $\approx$1065~nm, bottom: $\approx$1372~nm) of the extracted image cubes using the CHARIS-based pipeline with \textit{\textup{optimal extraction}} in hexagonal geometry (left panel) and resampled on a rectilinear grid (middle panel), and the extraction using the DRH-based Data Center reduction (right panel). Section~\ref{sec:extraction} describes the extraction process in detail.

Figure~\ref{fig:comparison1} shows the overall behavior of the different extraction approaches. In high-S/N channels, our pipeline extractions are visually similar to the DRH/DC extraction. The largest difference appears in the lowest-S/N channel. In the case of DRH, a stripe pattern is visible from the spectral extraction. This pattern gradually becomes more dominant as the S/N decreases, but is present in all channels at low amplitude. This pattern is not visible in our pipeline reductions.
In data sets with very high detector counts ($>$20,000), the crosstalk correction in all reductions creates a negative feature at the location of the brightest speckles at shorter wavelengths in the lowest S/N channels, especially at the location of satellite spots. This might indicate that the satellite spots, if strongly exposed, may not be perfectly reliable.

\subsection{Spectrum of the white dwarf companion HD~2133B}

Our first quantitative test of the new cube extraction pipeline uses the white dwarf HD~2113B.  The white dwarf has an effective temperature of $\approx$28,000\,K \citep{Burleigh1997} and orbits the F7 star \object{HD 2133} at a separation of about $0.\!\!''68$.
It is bright enough to be easily visible in individual exposures and wavelength channels of the coronagraphic sequence. Hot white dwarfs have relatively simple spectra on their Rayleigh-Jeans tails in the near-infrared, making them good candidates for understanding spectrophotometric calibrations.

We extracted data cubes of HD~2133B using the DRH pipeline, our CHARIS-based pipeline with optimal extraction and least-squares extraction.  In all cases, we resampled the data onto a rectilinear grid to facilitate postprocessing.  Because the white dwarf is located at the adaptive-optics correction radius, we still performed a simple postprocessing step to reduce speckle noise while at the same time extracting the contrast of the white dwarf relative to its main-sequence binary component. To do this, we ran the \texttt{TRAP} temporal detrending algorithm without any principle components, which means that only an offset was fit together with the forward model of the PSF for each pixel that was affected by signal of the white dwarf companion. This corresponds to a temporal version of classical ADI \citep{Marois2006}, except that we simultaneously fit a temporal forward model and an offset to each affected pixel using the known PSF instead of subtracting a median image from each frame. We performed this parameter-free extraction of the white dwarf contrast for all three reductions (optimal and least-squares extraction and DRH/DC).

\begin{figure*}
    \centering
    \includegraphics[width=\textwidth]{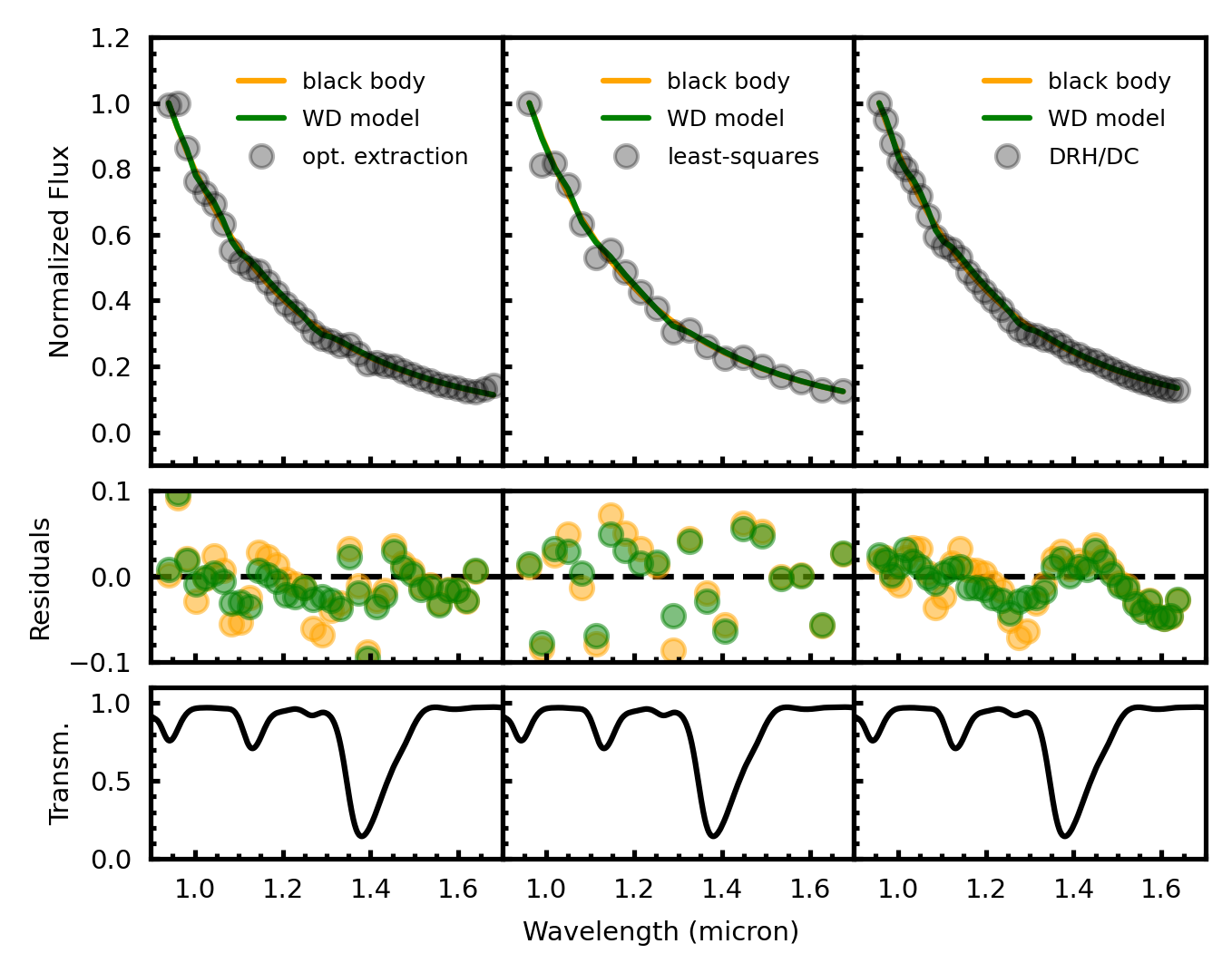}
    \caption{Extracted spectrum of the white dwarf HD 2133 B compared to a blackbody and a white dwarf model spectrum of the same temperature, 26,000~K. The extracted spectrum has a high S/N ($>$200) and follows the spectrum to within $\sim$5\%.}
    \label{fig:wd_spectrum}
\end{figure*}

We converted the measured contrast into a flux using a BT-NextGen model for the star. Based on comparing different combinations of model parameters for both the star and white dwarf taken from the literature with our data, we determined that the stellar parameters of $T_\textrm{eff}=6400\,$K, $\log\,g=4.3$, solar metallicity, and a white dwarf model of $T_\textrm{eff}=26000\,$K, $\log\,g=8.0$, provide a close fit to the data. The normalized fluxes were compared both to a simple blackbody of 26000~K and to a theoretical white dwarf spectrum from the \citet{Levenhagen2017} grid. Figure~\ref{fig:wd_spectrum} shows the extracted white dwarf spectra compared to models. As the S/N is very high (>200), the statistical uncertainties of the extracted contrast are small (not shown because they are smaller than the symbols); the residuals are dominated by other effects. We also show the average atmospheric transmission profile obtained for Cerro Paranal Observatory using \textsc{SkyCalc} \citep{SkyCalc2012} to provide context for small deviations from the expected spectral shape.

The two pipelines and the different extraction methods are all consistent with the model at a level of $\lesssim$5\%. As expected, the white dwarf model provides a slightly better match to the data than the blackbody because it includes the main absorption features of the white dwarf spectrum.
We computed two metrics for the three extractions: first the median absolute deviation of the residuals from the white dwarf model, 
\begin{equation}
\label{eq:mad}
    {\displaystyle \operatorname {MAD} =\operatorname {median} {(|\vec{x}|})},
\end{equation}
where $|\vec{x}|$ are the absolute values of the model residuals normalized to the model flux. Second, we computed the correlation between pairs of residual values at adjacent wavelengths.
Assuming that $\vec{x}$ has $N$ elements, we computed the neighbor correlation as 
\begin{equation}
\label{eq:corr}
    \begin{split}
        \psi &= \left(  \sum\limits_{i=1}^{N-1} (x_{i} \, x_{i+1}) + x_{N} \cdot x_{1} \right) \left( \sum\limits_{i=1}^{N} x_{i}^2 \right)^{-1}
    \end{split}
,\end{equation}
where the last term uses periodic boundary conditions to match the number of terms in the numerator and denominator. 
Table~\ref{tab:residual_metrics} summarizes the results of these two metrics from Eq.~\ref{eq:mad}~and~\ref{eq:corr} for the three extractions.

\begin{table}[t]
\caption[White dwarf model residual metrics]{White dwarf model residual metrics}
\centering
\begin{tabular}{l c c}
\hline\hline
Extraction & MAD & Correlation $\psi$ \\
\hline
optimal extraction & 1.9\% & 0.48 \\
least-squares extraction & 3.7\% & -0.05 \\
DRH/DC & 1.7\% & 0.84 \\
\hline
\end{tabular}
\label{tab:residual_metrics}
\end{table}
As our pipeline performs fewer interpolations (e.g., no post-hoc wavelength scaling, no anamorphism correction), the residuals are less spectrally correlated than in the DRH pipeline, and as expected from \citetalias{Brandt2017}, they are slightly anticorrelated for least-squares extraction. In the case of least-squares extraction, which effectively deconvolves the images with the line-spread function, the spectrum appears to scatter more (seen in the higher MAD value). However, this is an artifact that arises because the spectrum is less strongly smoothed. The residual scatter of the optimal extraction and DRH/DC reduction is comparable (difference of 0.2\%). All reductions show a high reliability of the spectral extraction.

We note that there are kinks in the residuals that largely correspond to the spectral channels that are most strongly affected by a change in atmospheric transmission. These are an indication that taking the atmospheric transmission in the model spectra into account becomes important in low spectral resolution spectroscopy. This is an independent problem that is not related to the spectral extraction of the IFS data itself, however.

\subsection{Spectrum of 51 Eridani b}

As a second test of our pipeline, we extracted and postprocessed data cubes of the directly imaged exoplanet 51~Eri~b \citep{Macintosh2015}. 
We used two coronagraphic datasets of \object{51 Eri} obtained by SPHERE IFS: the data from 2015 \citep{Samland2017}, and newer data from 2017 \citep{Brown2022}. We extracted data cubes using our pipeline, and for comparison, we used the extracted image cubes provided by the SPHERE Data Center using the ESO DRH pipeline. The DC by default temporally bins the data with a factor that avoids significant smearing in the IFS FoV. In these data sets, this results in 60s exposures, using a binning of four for 2015 and two in 2017.
We then extracted the planetary spectrum using the \texttt{TRAP} \citep{Samland2021} postprocessing algorithm with the same uniform default settings for all data sets. 

\begin{figure*}
    \centering
    \begin{tabular}{cc}
    \includegraphics[width=0.5\textwidth]{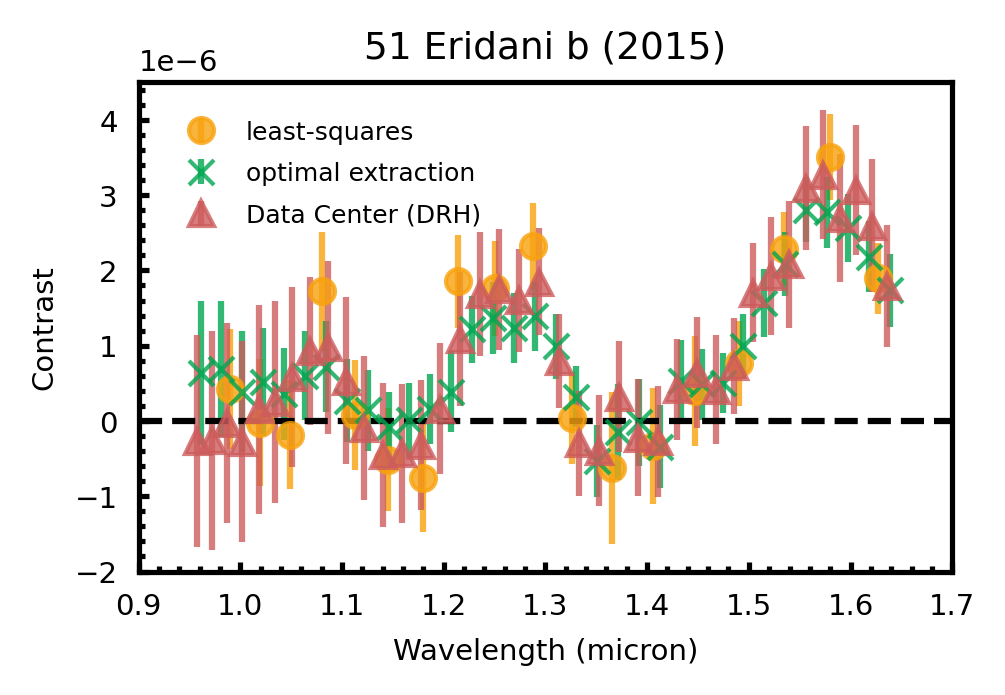} &
    \includegraphics[width=0.5\textwidth]{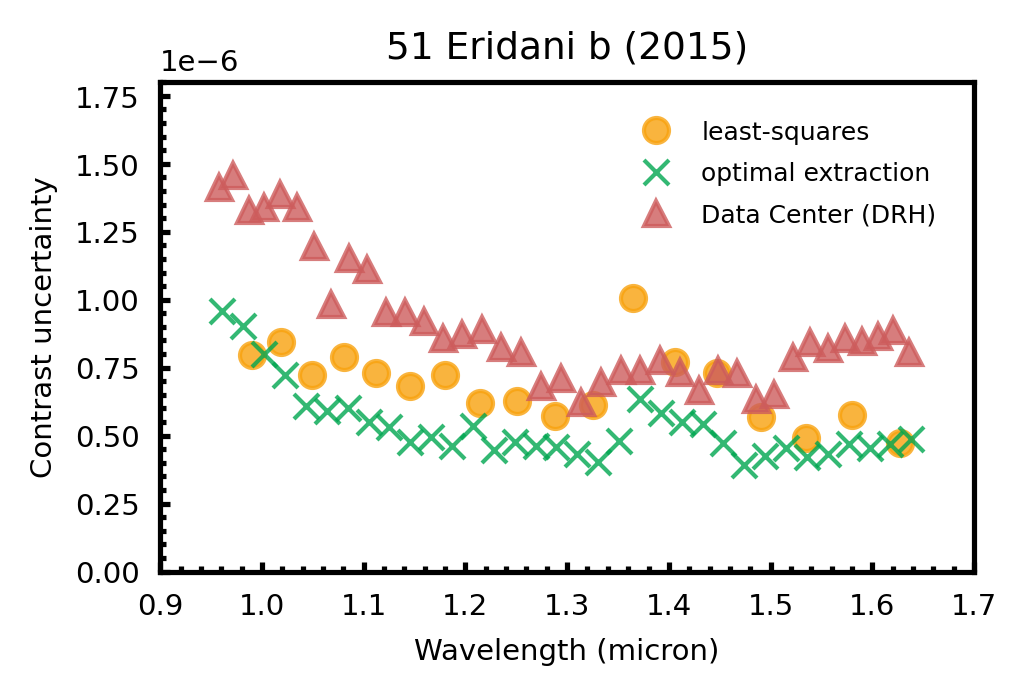} \\
    \includegraphics[width=0.5\textwidth]{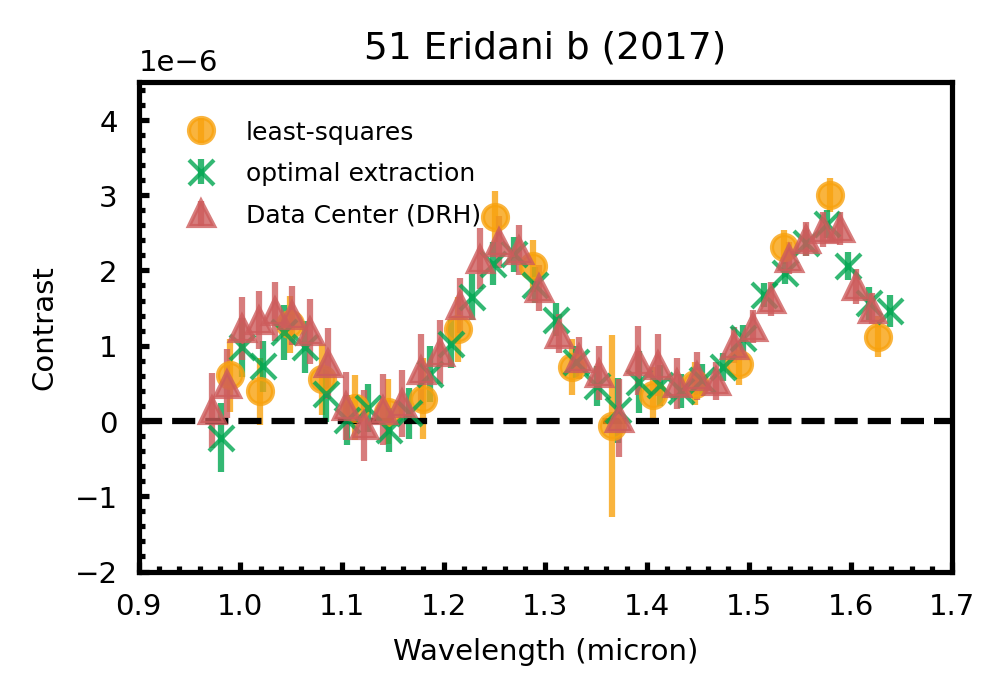} & 
    \includegraphics[width=0.5\textwidth]{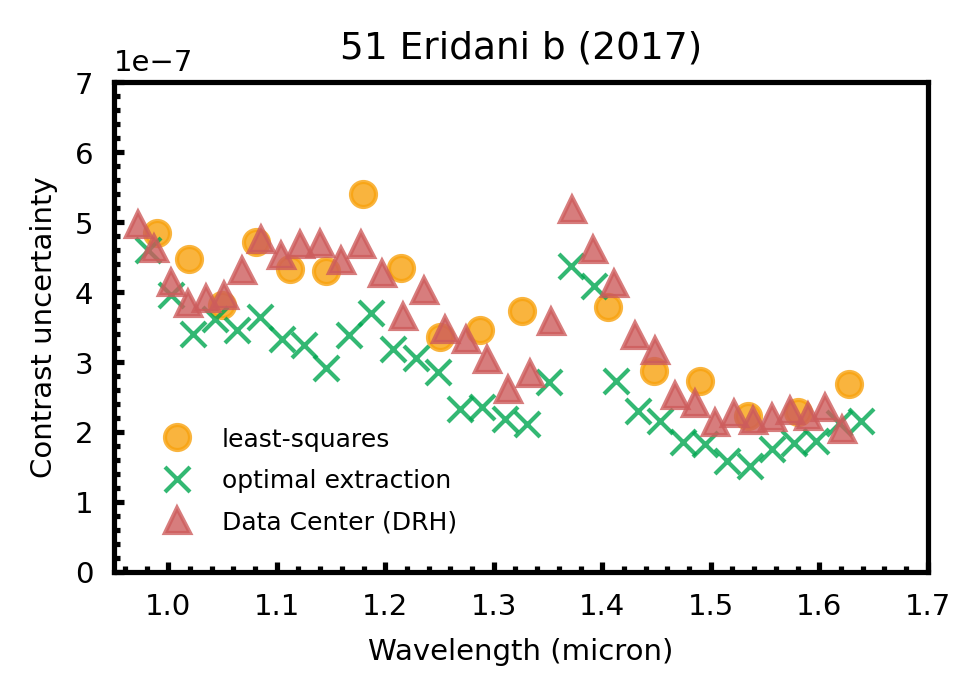}
    \end{tabular}
    \caption{Spectra extracted from 51~Eridani~b data using different spectral cube extraction methods but the same postprocessing. The top row shows data from 2015 taken in challenging conditions with a strong wind-driven halo, and the bottom row shows data from 2017 taken under excellent conditions. The left panels show the extracted planet spectra, and the right panels show only the uncertainties of the extracted spectrum for better clarity.}
    \label{fig:51Eri_spectra}
\end{figure*}

Figure~\ref{fig:51Eri_spectra} shows the resulting spectra and uncertainties for the 2015 and 2017 data sets and for each of the three cube-extraction approaches. The upper panels show the spectrum obtained from the 2015 data, the lower panels correspond to the 2017 data. The left panels show the spectra, and the right panels show the statistical uncertainties from the extraction for better visibility because the spectra overlap strongly. 

The statistical uncertainties in both data sets are smallest after postprocessing when the optimal extraction method of our pipeline is used. The least-squares method also reduces uncertainties in the extraction compared to DRH/DC in the 2015 data, but not as much in the cleaner 2017 data. However, it should be noted that least-squares performs a deconvolution with the line-spread function, therefore the spectral correlations are reduced and the spectral sampling is sparser, at the cost of larger scatter (less smoothed-out images). The direct comparison of the 1D uncertainties may therefore be misleading. On the other hand, optimal extraction is directly comparable to the DRH reduction, and there is a consistent reduction in the statistical uncertainties by a factor of $1.80^{+0.70}_{-0.67}$ for the 2015 data and $1.31^{+0.39}_{-0.37}$ for the 2017 data. These numbers refer to the median and 16--84 percentile range.

\begin{figure*}
    \centering
    \begin{tabular}{c}
    \includegraphics[width=\textwidth]{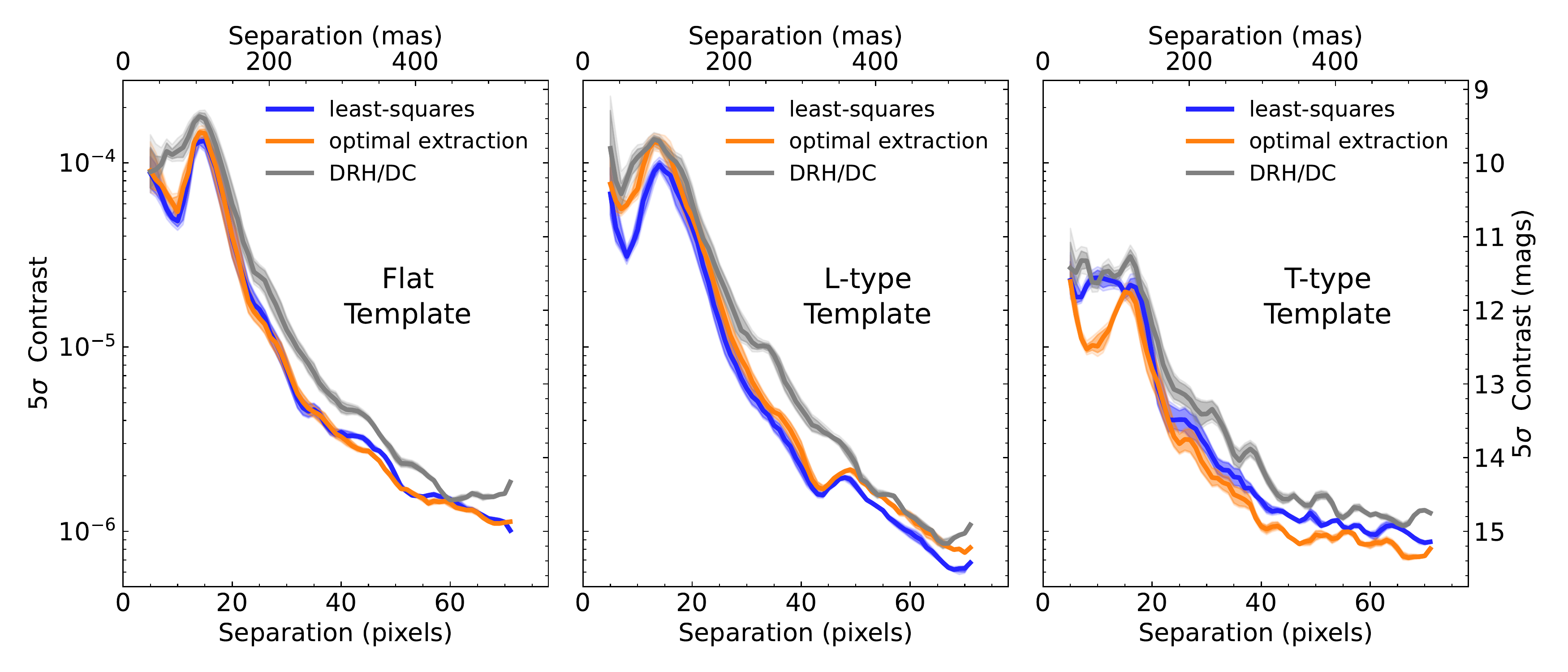} \\
    \includegraphics[width=\textwidth]{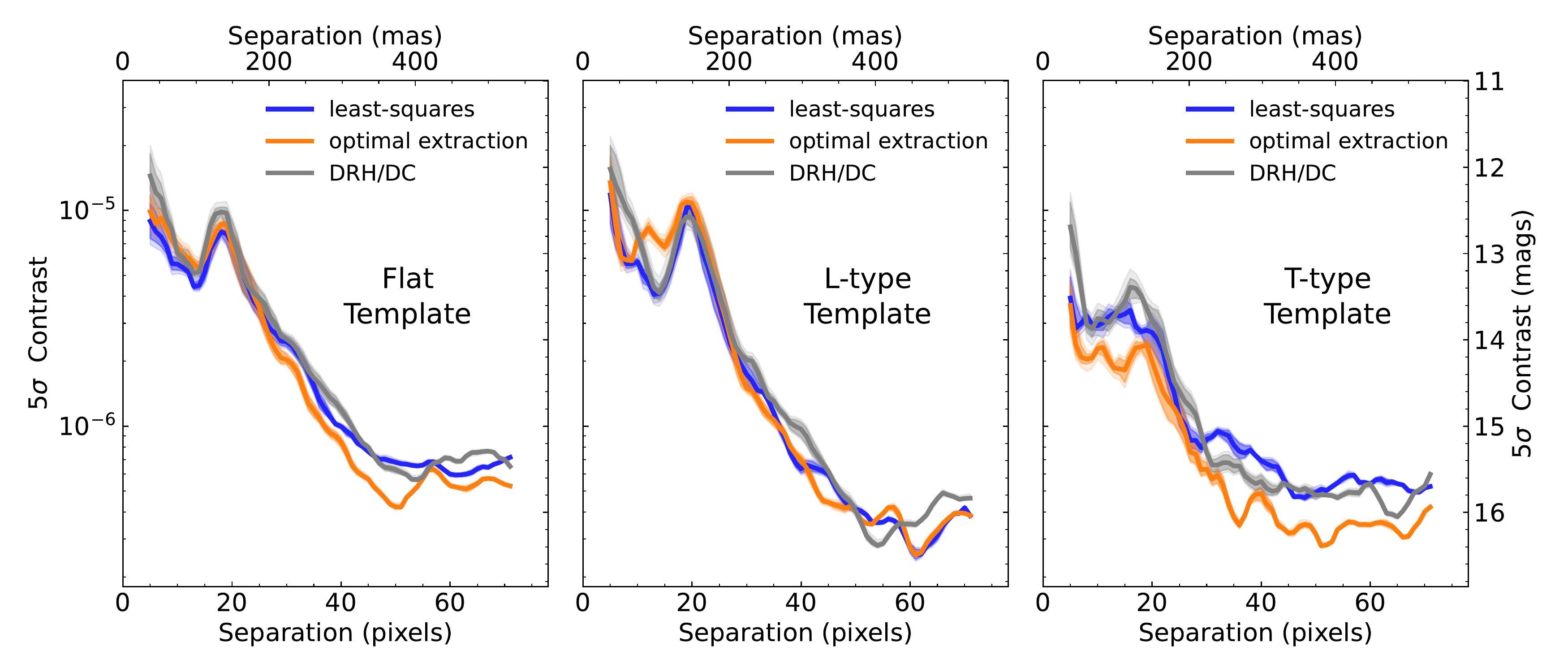} \\
    \end{tabular}
    \caption{Contrast curves for 51~Eri obtained from using \texttt{TRAP} on image cubes extracted using our pipeline and the DRH/DC pipeline. The upper row shows results for the 2015 data, and the lower row shows the results for the better 2017 data. The individual panels show different spectral templates of planetary companions that were used to combined the contrast curves over the IFS wavelength channels: flat contrast, which corresponds to a mean contrast over all channels, and L-type and T-type spectral contrast templates.}
    \label{fig:51Eri_2015_contrast}
\end{figure*}

In Fig.~\ref{fig:51Eri_2015_contrast} we show the corresponding contrast curves for these reductions. The contrast is given for different spectral templates (left: flat contrast, middle: L-type, and right: T-type). The least-squares and optimal extraction method of our pipeline provide comparable results and show a corresponding similar improvement compared to the DRH/DC reduction.

The contrast obtained with postprocessing of our pipeline-reduced images is consistently improved with respect to the DRH/DC. The strongest improvement is noticeable in the 2015 data using optimal extraction as was already seen in the extraction of the spectrum. The spectral template used to combine the wavelength channels does not impact these results significantly. As expected, we are most sensitive to T-type objects because they contain the most distinct spectral features compared to residual speckles.

\section{Usage and performance}
\label{sec:usage_performance}
The adapted CHARIS pipeline is publicly available on Github. It has been modularized such that all instrument-specific calibrations and parameters are moved into a class structure, such that it can easily be maintained and extended for future instruments if necessary.

The pipeline functions can either be called from Python or a command-line tool can be used to extract image cubes using a parameter file. The latter was the main way of interfacing with the original CHARIS pipeline. Calling the Python module directly is recommended for constructing an optimized workflow, however.

As with the previous CHARIS-pipeline, it only requires two steps to produce results: 1) `buildcal' generates the calibration files, the wavelength calibration and (oversampled) PSFlet models, from a monochromatic flat-field for a specific night containing the laser spots, and 2) `extractcube' extracts an image cube from raw image input and the calibration files.
The usage of the command-line tools is analogous to the previous version (\citetalias{Brandt2017}) for the calibration step. For the extraction step, the only difference is that three more parameters can be specified in the [Extract] category of the config-file: 1) \texttt{crosstalkscale} (float), which provides a continuous tuning of the crosstalk correction.  The default of 1.0 provides a full crosstalk correction (see \ref{sec:optimal_extraction}), and 0 provides no crosstalk correction. 
2) \texttt{dcxtalkcorrection} (boolean), if set \textit{True} tells the pipeline to use the same crosstalk correction as implemented in the Data Center and \citet{Vigan2020_pipeline} instead of our method, which is based on the fitted PSFlet model. This alternative crosstalk correction smooths the data with a modified Moffat-kernel and subtracts the smoothed image to remove large-scale structures from the unextracted raw images.
3) \texttt{individual\_dits} (boolean), if set \textit{True} every individual exposure (detector integration, dit) in the data cube is extracted separately. Otherwise, the cube is averaged and then extracted. We strongly recommend extracting all exposures for obtaining the highest temporal resolution, but this parameter can be set to \textit{False} for testing purposes in order to speed up the extraction.
For SPHERE IFS data, if \texttt{bgsub} is \textit{True} and no background images are provided, the pipeline uses the new principal component based fitting of the thermal background to perform the background subtraction.
When data are directly reduced from Python, the \textit{extractcube.getcube}-function can be called. We refer to the corresponding documentation for details of the required input parameters. An example script will be provided.

For non-coronagraphic PSF images taken with SPHERE IFS, the extraction may fail when \texttt{fitshift} is set to True. The reason for this is that in the case of SPHERE an overall subpixel shift of the spectra is determined from all lenslets. However, for non-coronagraphic PSF images only a very small fraction of lenslets are exposed to the light of the star making the determination unreliable and in some cases fail.

\paragraph{Performance}
The `buildcal' step is by far the most time-consuming part of a reduction. On a typical laptop and a single core, this can take about 90 minutes when the `upsample' option is switched on, which is recommended for best results. However, this step is only required once for any particular observation night, and the resulting calibration can be used for any of the targets that are observed on that date and can be further sped up by using multiple cores for an approximately linear decrease in computational time with the number of cores.

The extraction of a single exposure from a cube of raw images on one core takes less than three minutes on the highest setting, that is, using the oversampled calibrations and iterative fitting of the PSFlet models to the data to subtract crosstalk. This includes all the processing steps, including the identification and correction of bad hexagons based on an iterative comparison of each hexagon with its neighbours, and the subsequent resampling of the data on a rectilinear grid.

\section{Discussion and conclusion}
\label{sec:discussion_conclusion}
We presented a new open-source data-reduction pipeline for the SPHERE IFS instrument. Our pipeline is a candidate replacement for the current modified DRH pipeline that is the basis of all IFS data reduction. It makes use of new calibration data that were specifically obtained for the purpose of improving the IFS data reduction. The pipeline is straightforward to use for anyone with a laptop and a current Python distribution, but can also make use of heavy multiprocessing if a large amount of data is processed. We plan to include our pipeline into the SPHERE Data Center (DC) in the future to facilitate access for the community even more. Because of the significant improvement in data quality provided by our pipeline over the current version of the SPHERE-DC pipeline, the SPHERE-DC \citep{SPHEREDC} has started adapting our pipeline to its infrastructure. It will then gradually reprocess and publicly release all available public IFS data over the next few years. Our pipeline was also recently used in \citet{Franson2022} as the primary reduction for a newly discovered companion.

The public DRH has several important and known limitations that prevent its use for the most advanced science, such as limited accuracy in the wavelength calibration (up to 20~nm discrepancy), a grid pattern that is especially visible in low-S/N channels, and the complicated structure of the reduction recipes, which requires expert knowledge.
A user wishing to postprocess SPHERE data must therefore either contact the Data Center or use a Python wrapper for the DRH pipeline containing known fixes for some of the issues \citep{Vigan2020_pipeline}, such as correcting the wavelength calibration based on measurements of the position of the satellite spots. This poses other dangers, however; the transmission profile of the atmosphere will shift the effective wavelength measured for the satellite spot position in channels that show water or methane absorption. The grid pattern visible in the DRH reduction is particularly detrimental in the study of extended low-S/N structures such as  disks, and for obtaining reliable measurements in the water- or methane-affected spectral range of exoplanets and brown dwarfs. It also causes problems for sparse aperture masking data by introducing noise at specific spatial frequencies. The extracted data cubes are also often binned temporally and always contain multiple interpolations, both spectral and spatial.

The SPHERE IFS pipeline presented here, in contrast, inherently avoids these issues. 
Our pipeline introduces a number of improvements over the current Data Center product based on the ESO pipeline. Our pipeline improvements are enabled by new calibration data: monochromatic images of the individual calibration lasers taken through an integrating sphere.  These enable us to extract and reconstruct the monochromatic lenslet PSFs at each of the four calibration wavelengths and over all regions of the detector.  

The monochromatic lenslet PSFs enable fundamental improvements and new approaches to the cube reconstruction process. We can reconstruct the 2D shape of the spectra on the detector using a forward-modeling approach using our high-quality lenslet PSFs measurements. The model can be fit with a least-squares extraction that allows us to avoid interpolating the data spectrally. This forward model is also used to correct the crosstalk in a self-consistent fashion that does not rely on removing large-scale structures with a smoothing kernel.  
In addition to a least-squares extraction, we also enable optimal extraction using the measured lenslet PSF profiles, which is a straight  improvement compared to the aperture photometry approach implemented in the DRH.
We implemented a PCA-based approach to thermal background subtraction that makes taking sky frames unneccesary. It avoids introduction of additional read and photon noise, and persistence noise from sky frames taken after coronagraphic images.

Our pipeline allows the preservation of the original hexagon geometry of the images, which avoids the most aggressive spatial interpolation step in the data-reduction process. This opens the way for many possible improvements in postprocessing. However, in order to use traditional image-based postprocessing algorithms on hexagonal images, it becomes necessary to implement algorithms for registering and aligning (and potentially rotating) hexagon images. Furthermore, algorithms that work on image regions need to select these regions based on the real hexagon position instead of 2D array indices, which is often how the selection is implemented.
For \texttt{TRAP} \citep{Samland2021}, only this second step is necessary because the algorithm does not require the images to be aligned. The companion PSF forward model, however, still has to be shifted to the correct position. This might be simplified by using an analytic model trained on the unsatured PSF image, thereby avoiding this issue entirely.
An algorithm such as \texttt{TRAP} that does not require images to be aligned because it works directly on the time-series of pixels or spaxels can work on the completely uninterpolated spaxels, thereby reducing noise correlations significantly. This will further improve the attainable contrast, especially because our pipeline provides uncertainties for each spaxel of the extracted image that can be used as input for postprocessing.
The necessary changes have not yet been implemented into \texttt{TRAP}, but this is planned for the future. In general, the architecture of \texttt{TRAP} is such that the required changes are relatively minor (passing the position of each hexagon to the reduction area and reference area selection routine and adjusting the forward model of the companion PSF to the lenslet location in each frame of the ADI sequence).

Our SPHERE IFS pipeline is easy to use as it has only two steps: First, the calibration data are build from the \texttt{wavecal} file, a monochromatic measurement of the lenslet PSF using three to four lasers depending on the wavelength coverage. This step only requires one calibration file, all other calibrations are included with the pipeline. Second, the calibration data are used to extract the spectral cubes.
In conclusion, we presented a new pipeline that addresses multiple known issues, introduces several significant improvements, is accessible, and is easy to use, with much potential for further improvement of postprocessing in the future.



\begin{acknowledgements}
MS acknowledges support from the European Research Council under the Horizon 2020 Framework Program via the ERC Advanced Grant Origins 83 24 28.
This work has made use of the the SPHERE Data Centre, jointly operated by OSUG/IPAG (Grenoble), PYTHEAS/LAM/CESAM (Marseille), OCA/Lagrange (Nice), Observatoire de Paris/LESIA (Paris), and Observatoire de Lyon. 
This work is supported by the French National Research Agency in the framework of the Investissements d’Avenir program (ANR-15-IDEX-02), through the funding of the ``Origin of Life'' project of the Grenoble-Alpes University. This project has received funding from the European Research Council (ERC) under the European Union's Horizon 2020 research and innovation programme (COBREX; grant agreement n$^\circ$ 885593.
\end{acknowledgements}


    \bibliographystyle{aa} 
    \bibliography{bibliography} 

\begin{thebibliography}{61}
\expandafter\ifx\csname natexlab\endcsname\relax\def\natexlab#1{#1}\fi

\bibitem[{{Anderson} \& {King}(2000)}]{Anderson+King_2000}
{Anderson}, J. \& {King}, I.~R. 2000, \pasp, 112, 1360

\bibitem[{{Antichi} {et~al.}(2009){Antichi}, {Dohlen}, {Gratton}, {Mesa},
  {Claudi}, {Giro}, {Boccaletti}, {Mouillet}, {Puget}, \&
  {Beuzit}}]{Antichi2009}
{Antichi}, J., {Dohlen}, K., {Gratton}, R.~G., {et~al.} 2009, \apj, 695, 1042

\bibitem[{{Bailey}(2012)}]{Bailey2012}
{Bailey}, S. 2012, \pasp, 124, 1015

\bibitem[{{Berdeu} {et~al.}(2020){Berdeu}, {Soulez}, {Denis}, {Langlois}, \&
  {Thi{\'e}baut}}]{Berdeu+Soulez+Ferreol+etal_2020}
{Berdeu}, A., {Soulez}, F., {Denis}, L., {Langlois}, M., \& {Thi{\'e}baut},
  {\'E}. 2020, \aap, 635, A90

\bibitem[{{Beuzit} {et~al.}(2019){Beuzit}, {Vigan}, {Mouillet}, {Dohlen},
  {Gratton}, {Boccaletti}, {Sauvage}, {Schmid}, {Langlois}, {Petit},
  {Baruffolo}, {Feldt}, {Milli}, {Wahhaj}, {Abe}, {Anselmi}, {Antichi},
  {Barette}, {Baudrand}, {Baudoz}, {Bazzon}, {Bernardi}, {Blanchard}, {Brast},
  {Bruno}, {Buey}, {Carbillet}, {Carle}, {Cascone}, {Chapron}, {Charton},
  {Chauvin}, {Claudi}, {Costille}, {De Caprio}, {de Boer}, {Delboulb{\'e}},
  {Desidera}, {Dominik}, {Downing}, {Dupuis}, {Fabron}, {Fantinel}, {Farisato},
  {Feautrier}, {Fedrigo}, {Fusco}, {Gigan}, {Ginski}, {Girard}, {Giro},
  {Gisler}, {Gluck}, {Gry}, {Henning}, {Hubin}, {Hugot}, {Incorvaia}, {Jaquet},
  {Kasper}, {Lagadec}, {Lagrange}, {Le Coroller}, {Le Mignant}, {Le Ruyet},
  {Lessio}, {Lizon}, {Llored}, {Lundin}, {Madec}, {Magnard}, {Marteaud},
  {Martinez}, {Maurel}, {M{\'e}nard}, {Mesa}, {M{\"o}ller-Nilsson}, {Moulin},
  {Moutou}, {Orign{\'e}}, {Parisot}, {Pavlov}, {Perret}, {Pragt}, {Puget},
  {Rabou}, {Ramos}, {Reess}, {Rigal}, {Rochat}, {Roelfsema}, {Rousset}, {Roux},
  {Saisse}, {Salasnich}, {Santambrogio}, {Scuderi}, {Segransan}, {Sevin},
  {Siebenmorgen}, {Soenke}, {Stadler}, {Suarez}, {Tiph{\`e}ne}, {Turatto},
  {Udry}, {Vakili}, {Waters}, {Weber}, {Wildi}, {Zins}, \&
  {Zurlo}}]{Beuzit2019}
{Beuzit}, J.~L., {Vigan}, A., {Mouillet}, D., {et~al.} 2019, \aap, 631, A155

\bibitem[{{Boccaletti} {et~al.}(2015){Boccaletti}, {Thalmann}, {Lagrange},
  {Janson}, {Augereau}, {Schneider}, {Milli}, {Grady}, {Debes}, {Langlois},
  {Mouillet}, {Henning}, {Dominik}, {Maire}, {Beuzit}, {Carson}, {Dohlen},
  {Engler}, {Feldt}, {Fusco}, {Ginski}, {Girard}, {Hines}, {Kasper}, {Mawet},
  {M{\'e}nard}, {Meyer}, {Moutou}, {Olofsson}, {Rodigas}, {Sauvage},
  {Schlieder}, {Schmid}, {Turatto}, {Udry}, {Vakili}, {Vigan}, {Wahhaj}, \&
  {Wisniewski}}]{Boccaletti2015}
{Boccaletti}, A., {Thalmann}, C., {Lagrange}, A.-M., {et~al.} 2015, \nat, 526,
  230

\bibitem[{{Bohn} {et~al.}(2020){Bohn}, {Kenworthy}, {Ginski}, {Manara},
  {Pecaut}, {de Boer}, {Keller}, {Mamajek}, {Meshkat}, {Reggiani}, {Todorov},
  \& {Snik}}]{Bohn2020}
{Bohn}, A.~J., {Kenworthy}, M.~A., {Ginski}, C., {et~al.} 2020, \mnras, 492,
  431

\bibitem[{{Bonavita} {et~al.}(2022){Bonavita}, {Fontanive}, {Gratton},
  {Mu{\v{z}}i{\'c}}, {Desidera}, {Mesa}, {Biller}, {Scholz}, {Sozzetti}, \&
  {Squicciarini}}]{Bonavita2022}
{Bonavita}, M., {Fontanive}, C., {Gratton}, R., {et~al.} 2022, \mnras, 513,
  5588

\bibitem[{{Brandt} {et~al.}(2017){Brandt}, {Rizzo}, {Groff}, {Chilcote},
  {Greco}, {Kasdin}, {Limbach}, {Galvin}, {Loomis}, {Knapp}, {McElwain},
  {Jovanovic}, {Currie}, {Mede}, {Tamura}, {Takato}, \& {Hayashi}}]{Brandt2017}
{Brandt}, T.~D., {Rizzo}, M., {Groff}, T., {et~al.} 2017, Journal of
  Astronomical Telescopes, Instruments, and Systems, 3, 048002

\bibitem[{{Brown-Sevilla} {et~al.}(2022){Brown-Sevilla}, {Maire}, \&
  {Molli{\`e}re}}]{Brown2022}
{Brown-Sevilla}, S.~B., {Maire}, A.-L., \& {Molli{\`e}re}, P. 2022, \aap,
  submitted

\bibitem[{{Burleigh} {et~al.}(1997){Burleigh}, {Barstow}, \&
  {Fleming}}]{Burleigh1997}
{Burleigh}, M.~R., {Barstow}, M.~A., \& {Fleming}, T.~A. 1997, \mnras, 287, 381

\bibitem[{{Cantalloube} {et~al.}(2015){Cantalloube}, {Mouillet}, {Mugnier},
  {Milli}, {Absil}, {Gomez Gonzalez}, {Chauvin}, {Beuzit}, \&
  {Cornia}}]{Cantalloube2015}
{Cantalloube}, F., {Mouillet}, D., {Mugnier}, L.~M., {et~al.} 2015, \aap, 582,
  A89

\bibitem[{{Chauvin} {et~al.}(2017{\natexlab{a}}){Chauvin}, {Desidera},
  {Lagrange}, {Vigan}, {Feldt}, {Gratton}, {Langlois}, {Cheetham}, {Bonnefoy},
  \& {Meyer}}]{Chauvin2017}
{Chauvin}, G., {Desidera}, S., {Lagrange}, A.~M., {et~al.} 2017{\natexlab{a}},
  in SF2A-2017: Proceedings of the Annual meeting of the French Society of
  Astronomy and Astrophysics, ed. C.~{Reyl{\'e}}, P.~{Di Matteo}, F.~{Herpin},
  E.~{Lagadec}, A.~{Lan{\c{c}}on}, Z.~{Meliani}, \& F.~{Royer}, Di

\bibitem[{{Chauvin} {et~al.}(2017{\natexlab{b}}){Chauvin}, {Desidera},
  {Lagrange}, {Vigan}, {Gratton}, {Langlois}, {Bonnefoy}, {Beuzit}, {Feldt},
  {Mouillet}, {Meyer}, {Cheetham}, {Biller}, {Boccaletti}, {D'Orazi},
  {Galicher}, {Hagelberg}, {Maire}, {Mesa}, {Olofsson}, {Samland}, {Schmidt},
  {Sissa}, {Bonavita}, {Charnay}, {Cudel}, {Daemgen}, {Delorme},
  {Janin-Potiron}, {Janson}, {Keppler}, {Le Coroller}, {Ligi}, {Marleau},
  {Messina}, {Molli{\`e}re}, {Mordasini}, {M{\"u}ller}, {Peretti}, {Perrot},
  {Rodet}, {Rouan}, {Zurlo}, {Dominik}, {Henning}, {Menard}, {Schmid},
  {Turatto}, {Udry}, {Vakili}, {Abe}, {Antichi}, {Baruffolo}, {Baudoz},
  {Baudrand}, {Blanchard}, {Bazzon}, {Buey}, {Carbillet}, {Carle}, {Charton},
  {Cascone}, {Claudi}, {Costille}, {Deboulbe}, {De Caprio}, {Dohlen},
  {Fantinel}, {Feautrier}, {Fusco}, {Gigan}, {Giro}, {Gisler}, {Gluck},
  {Hubin}, {Hugot}, {Jaquet}, {Kasper}, {Madec}, {Magnard}, {Martinez},
  {Maurel}, {Le Mignant}, {M{\"o}ller-Nilsson}, {Llored}, {Moulin},
  {Orign{\'e}}, {Pavlov}, {Perret}, {Petit}, {Pragt}, {Puget}, {Rabou},
  {Ramos}, {Rigal}, {Rochat}, {Roelfsema}, {Rousset}, {Roux}, {Salasnich},
  {Sauvage}, {Sevin}, {Soenke}, {Stadler}, {Suarez}, {Weber}, {Wildi},
  {Antoniucci}, {Augereau}, {Baudino}, {Brandner}, {Engler}, {Girard}, {Gry},
  {Kral}, {Kopytova}, {Lagadec}, {Milli}, {Moutou}, {Schlieder},
  {Szul{\'a}gyi}, {Thalmann}, \& {Wahhaj}}]{Chauvin2017_HIP65426}
{Chauvin}, G., {Desidera}, S., {Lagrange}, A.~M., {et~al.} 2017{\natexlab{b}},
  \aap, 605, L9

\bibitem[{{Chauvin} {et~al.}(2018){Chauvin}, {Gratton}, {Bonnefoy}, {Lagrange},
  {de Boer}, {Vigan}, {Beust}, {Lazzoni}, {Boccaletti}, {Galicher}, {Desidera},
  {Delorme}, {Keppler}, {Lannier}, {Maire}, {Mesa}, {Meunier}, {Kral},
  {Henning}, {Menard}, {Moor}, {Avenhaus}, {Bazzon}, {Janson}, {Beuzit},
  {Bhowmik}, {Bonavita}, {Borgniet}, {Brandner}, {Cheetham}, {Cudel}, {Feldt},
  {Fontanive}, {Ginski}, {Hagelberg}, {Janin-Potiron}, {Lagadec}, {Langlois},
  {Le Coroller}, {Messina}, {Meyer}, {Mouillet}, {Peretti}, {Perrot}, {Rodet},
  {Samland}, {Sissa}, {Olofsson}, {Salter}, {Schmidt}, {Zurlo}, {Milli}, {van
  Boekel}, {Quanz}, {Feautrier}, {Le Mignant}, {Perret}, {Ramos}, \&
  {Rochat}}]{Chauvin2018_HD95086}
{Chauvin}, G., {Gratton}, R., {Bonnefoy}, M., {et~al.} 2018, \aap, 617, A76

\bibitem[{{Cheetham} {et~al.}(2019){Cheetham}, {Samland}, {Brems}, {Launhardt},
  {Chauvin}, {S{\'e}gransan}, {Henning}, {Quirrenbach}, {Avenhaus}, {Cugno},
  {Girard}, {Godoy}, {Kennedy}, {Maire}, {Metchev}, {M{\"u}ller}, {Musso
  Barcucci}, {Olofsson}, {Pepe}, {Quanz}, {Queloz}, {Reffert}, {Rickman}, {van
  Boekel}, {Boccaletti}, {Bonnefoy}, {Cantalloube}, {Charnay}, {Delorme},
  {Janson}, {Keppler}, {Lagrange}, {Langlois}, {Lazzoni}, {Menard}, {Mesa},
  {Meyer}, {Schmidt}, {Sissa}, {Udry}, \& {Zurlo}}]{Cheetham2019}
{Cheetham}, A.~C., {Samland}, M., {Brems}, S.~S., {et~al.} 2019, \aap, 622, A80

\bibitem[{{Claudi} {et~al.}(2008){Claudi}, {Turatto}, {Gratton}, {Antichi},
  {Bonavita}, {Bruno}, {Cascone}, {De Caprio}, {Desidera}, {Giro}, {Mesa},
  {Scuderi}, {Dohlen}, {Beuzit}, \& {Puget}}]{Claudi2008}
{Claudi}, R.~U., {Turatto}, M., {Gratton}, R.~G., {et~al.} 2008, in \procspie,
  Vol. 7014, Ground-based and Airborne Instrumentation for Astronomy II, 70143E

\bibitem[{{Couder}(1949)}]{Couder1949}
{Couder}, A. 1949, L'Astronomie, 63, 253

\bibitem[{{Dahlqvist} {et~al.}(2020){Dahlqvist}, {Cantalloube}, \&
  {Absil}}]{Dahlqvist2020}
{Dahlqvist}, C.~H., {Cantalloube}, F., \& {Absil}, O. 2020, \aap, 633, A95

\bibitem[{{Delorme} {et~al.}(2017){Delorme}, {Meunier}, {Albert}, {Lagadec},
  {Le Coroller}, {Galicher}, {Mouillet}, {Boccaletti}, {Mesa}, {Meunier},
  {Beuzit}, {Lagrange}, {Chauvin}, {Sapone}, {Langlois}, {Maire},
  {Montarg{\`e}s}, {Gratton}, {Vigan}, \& {Surace}}]{SPHEREDC}
{Delorme}, P., {Meunier}, N., {Albert}, D., {et~al.} 2017, in SF2A-2017:
  Proceedings of the Annual meeting of the French Society of Astronomy and
  Astrophysics, ed. C.~{Reyl{\'e}}, P.~{Di Matteo}, F.~{Herpin}, E.~{Lagadec},
  A.~{Lan{\c{c}}on}, Z.~{Meliani}, \& F.~{Royer}, Di

\bibitem[{{Dohlen} {et~al.}(2008){Dohlen}, {Langlois}, {Saisse}, {Hill},
  {Origne}, {Jacquet}, {Fabron}, {Blanc}, {Llored}, {Carle}, {Moutou}, {Vigan},
  {Boccaletti}, {Carbillet}, {Mouillet}, \& {Beuzit}}]{Dohlen2008}
{Dohlen}, K., {Langlois}, M., {Saisse}, M., {et~al.} 2008, in \procspie, Vol.
  7014, Ground-based and Airborne Instrumentation for Astronomy II, 70143L

\bibitem[{{Franson} {et~al.}(2022){Franson}, {Bowler}, \&
  {Bonavita}}]{Franson2022}
{Franson}, K., {Bowler}, B.~P., \& {Bonavita}, M. 2022, \apj, submitted

\bibitem[{{Galicher} {et~al.}(2018){Galicher}, {Boccaletti}, {Mesa}, {Delorme},
  {Gratton}, {Langlois}, {Lagrange}, {Maire}, {Le Coroller}, {Chauvin},
  {Biller}, {Cantalloube}, {Janson}, {Lagadec}, {Meunier}, {Vigan},
  {Hagelberg}, {Bonnefoy}, {Zurlo}, {Rocha}, {Maurel}, {Jaquet}, {Buey}, \&
  {Weber}}]{specal2018}
{Galicher}, R., {Boccaletti}, A., {Mesa}, D., {et~al.} 2018, \aap, 615, A92

\bibitem[{{Gomez Gonzalez} {et~al.}(2016){Gomez Gonzalez}, {Wertz},
  {Christiaens}, {Absil}, \& {Mawet}}]{VIP2016}
{Gomez Gonzalez}, C.~A., {Wertz}, O., {Christiaens}, V., {Absil}, O., \&
  {Mawet}, D. 2016, {VIP: Vortex Image Processing pipeline for high-contrast
  direct imaging of exoplanets}, Astrophysics Source Code Library, record
  ascl:1603.003

\bibitem[{{Groff} {et~al.}(2016){Groff}, {Chilcote}, {Kasdin}, {Galvin},
  {Loomis}, {Carr}, {Brandt}, {Knapp}, {Limbach}, {Guyon}, {Jovanovic},
  {McElwain}, {Takato}, \& {Hayashi}}]{Groff2016}
{Groff}, T.~D., {Chilcote}, J., {Kasdin}, N.~J., {et~al.} 2016, in \procspie,
  Vol. 9908, Ground-based and Airborne Instrumentation for Astronomy VI, 99080O

\bibitem[{{Groff} {et~al.}(2015){Groff}, {Kasdin}, {Limbach}, {Galvin}, {Carr},
  {Knapp}, {Brandt}, {Loomis}, {Jarosik}, {Mede}, {McElwain}, {Leviton},
  {Miller}, {Quijada}, {Guyon}, {Jovanovic}, {Takato}, \&
  {Hayashi}}]{Groff2015}
{Groff}, T.~D., {Kasdin}, N.~J., {Limbach}, M.~A., {et~al.} 2015, in \procspie,
  Vol. 9605, Techniques and Instrumentation for Detection of Exoplanets VII,
  96051C

\bibitem[{{Guyon}(2005)}]{Guyon2005ApJ}
{Guyon}, O. 2005, \apj, 629, 592

\bibitem[{{Guyon} {et~al.}(2005){Guyon}, {Pluzhnik}, {Woodruff}, {Ridgway},
  {Galicher}, {Martinache}, \& {Blain}}]{Guyon2005AAS}
{Guyon}, O., {Pluzhnik}, E., {Woodruff}, R., {et~al.} 2005, in American
  Astronomical Society Meeting Abstracts, Vol. 207, American Astronomical
  Society Meeting Abstracts, 191.04

\bibitem[{{Hanu{\v{s}}} {et~al.}(2017){Hanu{\v{s}}}, {Marchis}, {Viikinkoski},
  {Yang}, \& {Kaasalainen}}]{Hanus2017}
{Hanu{\v{s}}}, J., {Marchis}, F., {Viikinkoski}, M., {Yang}, B., \&
  {Kaasalainen}, M. 2017, \aap, 599, A36

\bibitem[{{Hoeijmakers} {et~al.}(2018){Hoeijmakers}, {Schwarz}, {Snellen}, {de
  Kok}, {Bonnefoy}, {Chauvin}, {Lagrange}, \& {Girard}}]{Hoeijmakers2018}
{Hoeijmakers}, H.~J., {Schwarz}, H., {Snellen}, I.~A.~G., {et~al.} 2018, \aap,
  617, A144

\bibitem[{{Horne}(1986)}]{Horne1986}
{Horne}, K. 1986, \pasp, 98, 609

\bibitem[{{Janson} {et~al.}(2021){Janson}, {Squicciarini}, {Delorme},
  {Gratton}, {Bonnefoy}, {Reffert}, {Mamajek}, {Eriksson}, {Vigan}, {Langlois},
  {Engler}, {Chauvin}, {Desidera}, {Mayer}, {Marleau}, {Bohn}, {Samland},
  {Meyer}, {d'Orazi}, {Henning}, {Quanz}, {Kenworthy}, \&
  {Carson}}]{Janson2021a}
{Janson}, M., {Squicciarini}, V., {Delorme}, P., {et~al.} 2021, \aap, 646, A164

\bibitem[{{Jovanovic} {et~al.}(2015){Jovanovic}, {Martinache}, {Guyon},
  {Clergeon}, {Singh}, {Kudo}, {Garrel}, {Newman}, {Doughty}, {Lozi}, {Males},
  {Minowa}, {Hayano}, {Takato}, {Morino}, {Kuhn}, {Serabyn}, {Norris},
  {Tuthill}, {Schworer}, {Stewart}, {Close}, {Huby}, {Perrin}, {Lacour},
  {Gauchet}, {Vievard}, {Murakami}, {Oshiyama}, {Baba}, {Matsuo}, {Nishikawa},
  {Tamura}, {Lai}, {Marchis}, {Duchene}, {Kotani}, \&
  {Woillez}}]{Jovanovic2015b}
{Jovanovic}, N., {Martinache}, F., {Guyon}, O., {et~al.} 2015, \pasp, 127, 890

\bibitem[{{Keppler} {et~al.}(2018){Keppler}, {Benisty}, {M{\"u}ller},
  {Henning}, {van Boekel}, {Cantalloube}, {Ginski}, {van Holstein}, {Maire},
  {Pohl}, {Samland}, {Avenhaus}, {Baudino}, {Boccaletti}, {de Boer},
  {Bonnefoy}, {Chauvin}, {Desidera}, {Langlois}, {Lazzoni}, {Marleau},
  {Mordasini}, {Pawellek}, {Stolker}, {Vigan}, {Zurlo}, {Birnstiel},
  {Brandner}, {Feldt}, {Flock}, {Girard}, {Gratton}, {Hagelberg}, {Isella},
  {Janson}, {Juhasz}, {Kemmer}, {Kral}, {Lagrange}, {Launhardt}, {Matter},
  {M{\'e}nard}, {Milli}, {Molli{\`e}re}, {Olofsson}, {P{\'e}rez}, {Pinilla},
  {Pinte}, {Quanz}, {Schmidt}, {Udry}, {Wahhaj}, {Williams}, {Buenzli},
  {Cudel}, {Dominik}, {Galicher}, {Kasper}, {Lannier}, {Mesa}, {Mouillet},
  {Peretti}, {Perrot}, {Salter}, {Sissa}, {Wildi}, {Abe}, {Antichi},
  {Augereau}, {Baruffolo}, {Baudoz}, {Bazzon}, {Beuzit}, {Blanchard}, {Brems},
  {Buey}, {De Caprio}, {Carbillet}, {Carle}, {Cascone}, {Cheetham}, {Claudi},
  {Costille}, {Delboulb{\'e}}, {Dohlen}, {Fantinel}, {Feautrier}, {Fusco},
  {Giro}, {Gluck}, {Gry}, {Hubin}, {Hugot}, {Jaquet}, {Le Mignant}, {Llored},
  {Madec}, {Magnard}, {Martinez}, {Maurel}, {Meyer}, {M{\"o}ller-Nilsson},
  {Moulin}, {Mugnier}, {Orign{\'e}}, {Pavlov}, {Perret}, {Petit}, {Pragt},
  {Puget}, {Rabou}, {Ramos}, {Rigal}, {Rochat}, {Roelfsema}, {Rousset}, {Roux},
  {Salasnich}, {Sauvage}, {Sevin}, {Soenke}, {Stadler}, {Suarez}, {Turatto}, \&
  {Weber}}]{Keppler2018}
{Keppler}, M., {Benisty}, M., {M{\"u}ller}, A., {et~al.} 2018, \aap, 617, A44

\bibitem[{{Levenhagen} {et~al.}(2017){Levenhagen}, {Diaz}, {Coelho}, \&
  {Hubeny}}]{Levenhagen2017}
{Levenhagen}, R.~S., {Diaz}, M.~P., {Coelho}, P. R.~T., \& {Hubeny}, I. 2017,
  \apjs, 231, 1

\bibitem[{{Macintosh} {et~al.}(2015){Macintosh}, {Graham}, {Barman}, {De Rosa},
  {Konopacky}, {Marley}, {Marois}, {Nielsen}, {Pueyo}, {Rajan}, {Rameau},
  {Saumon}, {Wang}, {Patience}, {Ammons}, {Arriaga}, {Artigau}, {Beckwith},
  {Brewster}, {Bruzzone}, {Bulger}, {Burningham}, {Burrows}, {Chen}, {Chiang},
  {Chilcote}, {Dawson}, {Dong}, {Doyon}, {Draper}, {Duch{\^e}ne}, {Esposito},
  {Fabrycky}, {Fitzgerald}, {Follette}, {Fortney}, {Gerard}, {Goodsell},
  {Greenbaum}, {Hibon}, {Hinkley}, {Cotten}, {Hung}, {Ingraham},
  {Johnson-Groh}, {Kalas}, {Lafreniere}, {Larkin}, {Lee}, {Line}, {Long},
  {Maire}, {Marchis}, {Matthews}, {Max}, {Metchev}, {Millar-Blanchaer},
  {Mittal}, {Morley}, {Morzinski}, {Murray-Clay}, {Oppenheimer}, {Palmer},
  {Patel}, {Perrin}, {Poyneer}, {Rafikov}, {Rantakyr{\"o}}, {Rice}, {Rojo},
  {Rudy}, {Ruffio}, {Ruiz}, {Sadakuni}, {Saddlemyer}, {Salama}, {Savransky},
  {Schneider}, {Sivaramakrishnan}, {Song}, {Soummer}, {Thomas}, {Vasisht},
  {Wallace}, {Ward-Duong}, {Wiktorowicz}, {Wolff}, \&
  {Zuckerman}}]{Macintosh2015}
{Macintosh}, B., {Graham}, J.~R., {Barman}, T., {et~al.} 2015, Science, 350, 64

\bibitem[{{Macintosh} {et~al.}(2014){Macintosh}, {Graham}, {Ingraham},
  {Konopacky}, {Marois}, {Perrin}, {Poyneer}, {Bauman}, {Barman}, {Burrows},
  {Cardwell}, {Chilcote}, {De Rosa}, {Dillon}, {Doyon}, {Dunn}, {Erikson},
  {Fitzgerald}, {Gavel}, {Goodsell}, {Hartung}, {Hibon}, {Kalas}, {Larkin},
  {Maire}, {Marchis}, {Marley}, {McBride}, {Millar-Blanchaer}, {Morzinski},
  {Norton}, {Oppenheimer}, {Palmer}, {Patience}, {Pueyo}, {Rantakyro},
  {Sadakuni}, {Saddlemyer}, {Savransky}, {Serio}, {Soummer},
  {Sivaramakrishnan}, {Song}, {Thomas}, {Wallace}, {Wiktorowicz}, \&
  {Wolff}}]{Macintosh2014}
{Macintosh}, B., {Graham}, J.~R., {Ingraham}, P., {et~al.} 2014, Proceedings of
  the National Academy of Science, 111, 12661

\bibitem[{{Maire} {et~al.}(2020){Maire}, {Baudino}, {Desidera}, {Messina},
  {Brandner}, {Godoy}, {Cantalloube}, {Galicher}, {Bonnefoy}, {Hagelberg},
  {Olofsson}, {Absil}, {Chauvin}, {Henning}, \& {Langlois}}]{Maire2020_HD72946}
{Maire}, A.~L., {Baudino}, J.~L., {Desidera}, S., {et~al.} 2020, \aap, 633, L2

\bibitem[{{Maire} {et~al.}(2016){Maire}, {Langlois}, {Dohlen}, {Lagrange},
  {Gratton}, {Chauvin}, {Desidera}, {Girard}, {Milli}, {Vigan}, {Zins},
  {Delorme}, {Beuzit}, {Claudi}, {Feldt}, {Mouillet}, {Puget}, {Turatto}, \&
  {Wildi}}]{Maire2016b}
{Maire}, A.-L., {Langlois}, M., {Dohlen}, K., {et~al.} 2016, in SPIE Conf.
  Ser., Vol. 9908, 990834

\bibitem[{{Marois} {et~al.}(2006){Marois}, {Lafreni{\`e}re}, {Doyon},
  {Macintosh}, \& {Nadeau}}]{Marois2006}
{Marois}, C., {Lafreni{\`e}re}, D., {Doyon}, R., {Macintosh}, B., \& {Nadeau},
  D. 2006, \apj, 641, 556

\bibitem[{{Mesa} {et~al.}(2015){Mesa}, {Gratton}, {Zurlo}, {Vigan}, {Claudi},
  {Alberi}, {Antichi}, {Baruffolo}, {Beuzit}, {Boccaletti}, {Bonnefoy},
  {Costille}, {Desidera}, {Dohlen}, {Fantinel}, {Feldt}, {Fusco}, {Giro},
  {Henning}, {Kasper}, {Langlois}, {Maire}, {Martinez}, {Moeller-Nilsson},
  {Mouillet}, {Moutou}, {Pavlov}, {Puget}, {Salasnich}, {Sauvage}, {Sissa},
  {Turatto}, {Udry}, {Vakili}, {Waters}, \& {Wildi}}]{Mesa2015}
{Mesa}, D., {Gratton}, R., {Zurlo}, A., {et~al.} 2015, \aap, 576, A121

\bibitem[{{Milli} {et~al.}(2017{\natexlab{a}}){Milli}, {Hibon}, {Christiaens},
  {Choquet}, {Bonnefoy}, {Kennedy}, {Wyatt}, {Absil}, {G{\'o}mez Gonz{\'a}lez},
  {del Burgo}, {Matr{\`a}}, {Augereau}, {Boccaletti}, {Delacroix}, {Ertel},
  {Dent}, {Forsberg}, {Fusco}, {Girard}, {Habraken}, {Huby}, {Karlsson},
  {Lagrange}, {Mawet}, {Mouillet}, {Perrin}, {Pinte}, {Pueyo}, {Reyes},
  {Soummer}, {Surdej}, {Tarricq}, \& {Wahhaj}}]{Milli2017_HD206893}
{Milli}, J., {Hibon}, P., {Christiaens}, V., {et~al.} 2017{\natexlab{a}}, \aap,
  597, L2

\bibitem[{{Milli} {et~al.}(2018){Milli}, {Kasper}, {Bourget}, {Pannetier},
  {Mouillet}, {Sauvage}, {Reyes}, {Fusco}, {Cantalloube}, {Tristam}, {Wahhaj},
  {Beuzit}, {Girard}, {Mawet}, {Telle}, {Vigan}, \&
  {N'Diaye}}]{Milli2018SPIE_Wind}
{Milli}, J., {Kasper}, M., {Bourget}, P., {et~al.} 2018, in Society of
  Photo-Optical Instrumentation Engineers (SPIE) Conference Series, Vol. 10703,
  Adaptive Optics Systems VI, ed. L.~M. {Close}, L.~{Schreiber}, \&
  D.~{Schmidt}, 107032A

\bibitem[{{Milli} {et~al.}(2017{\natexlab{b}}){Milli}, {Vigan}, {Mouillet},
  {Lagrange}, {Augereau}, {Pinte}, {Mawet}, {Schmid}, {Boccaletti},
  {Matr{\`a}}, {Kral}, {Ertel}, {Chauvin}, {Bazzon}, {M{\'e}nard}, {Beuzit},
  {Thalmann}, {Dominik}, {Feldt}, {Henning}, {Min}, {Girard}, {Galicher},
  {Bonnefoy}, {Fusco}, {de Boer}, {Janson}, {Maire}, {Mesa}, {Schlieder}, \&
  {Sphere Consortium}}]{Milli2017}
{Milli}, J., {Vigan}, A., {Mouillet}, D., {et~al.} 2017{\natexlab{b}}, \aap,
  599, A108

\bibitem[{{M{\"u}ller} {et~al.}(2018){M{\"u}ller}, {Keppler}, {Henning},
  {Samland}, {Chauvin}, {Beust}, {Maire}, {Molaverdikhani}, {van Boekel},
  {Benisty}, {Boccaletti}, {Bonnefoy}, {Cantalloube}, {Charnay}, {Baudino},
  {Gennaro}, {Long}, {Cheetham}, {Desidera}, {Feldt}, {Fusco}, {Girard},
  {Gratton}, {Hagelberg}, {Janson}, {Lagrange}, {Langlois}, {Lazzoni}, {Ligi},
  {M{\'e}nard}, {Mesa}, {Meyer}, {Molli{\`e}re}, {Mordasini}, {Moulin},
  {Pavlov}, {Pawellek}, {Quanz}, {Ramos}, {Rouan}, {Sissa}, {Stadler}, {Vigan},
  {Wahhaj}, {Weber}, \& {Zurlo}}]{Mueller2018}
{M{\"u}ller}, A., {Keppler}, M., {Henning}, T., {et~al.} 2018, \aap, 617, L2

\bibitem[{{Noll} {et~al.}(2012){Noll}, {Kausch}, {Barden}, {Jones}, {Szyszka},
  {Kimeswenger}, \& {Vinther}}]{SkyCalc2012}
{Noll}, S., {Kausch}, W., {Barden}, M., {et~al.} 2012, \aap, 543, A92

\bibitem[{{Pavlov} {et~al.}(2008){Pavlov}, {M{\"o}ller-Nilsson}, {Feldt},
  {Henning}, {Beuzit}, \& {Mouillet}}]{Pavlov2008}
{Pavlov}, A., {M{\"o}ller-Nilsson}, O., {Feldt}, M., {et~al.} 2008, in
  \procspie, Vol. 7019, Advanced Software and Control for Astronomy II, 701939

\bibitem[{{Perrin} {et~al.}(2014){Perrin}, {Maire}, {Ingraham}, {Savransky},
  {Millar-Blanchaer}, {Wolff}, {Ruffio}, {Wang}, {Draper}, {Sadakuni},
  {Marois}, {Rajan}, {Fitzgerald}, {Macintosh}, {Graham}, {Doyon}, {Larkin},
  {Chilcote}, {Goodsell}, {Palmer}, {Labrie}, {Beaulieu}, {De Rosa},
  {Greenbaum}, {Hartung}, {Hibon}, {Konopacky}, {Lafreniere}, {Lavigne},
  {Marchis}, {Patience}, {Pueyo}, {Rantakyr{\"o}}, {Soummer},
  {Sivaramakrishnan}, {Thomas}, {Ward-Duong}, \& {Wiktorowicz}}]{Perrin2014}
{Perrin}, M.~D., {Maire}, J., {Ingraham}, P., {et~al.} 2014, in Society of
  Photo-Optical Instrumentation Engineers (SPIE) Conference Series, Vol. 9147,
  Ground-based and Airborne Instrumentation for Astronomy V, ed. S.~K.
  {Ramsay}, I.~S. {McLean}, \& H.~{Takami}, 91473J

\bibitem[{{Petit} {et~al.}(2012){Petit}, {Sauvage}, {Sevin}, {Costille},
  {Fusco}, {Baudoz}, {Beuzit}, {Buey}, {Charton}, {Dohlen}, {Feautrier},
  {Fedrigo}, {Gach}, {Hubin}, {Hugot}, {Kasper}, {Mouillet}, {Perret}, {Puget},
  {Sinquin}, {Soenke}, {Suarez}, \& {Wildi}}]{Petit2012}
{Petit}, C., {Sauvage}, J.~F., {Sevin}, A., {et~al.} 2012, in Society of
  Photo-Optical Instrumentation Engineers (SPIE) Conference Series, Vol. 8447,
  Adaptive Optics Systems III, 84471Z

\bibitem[{{Racine} {et~al.}(1999){Racine}, {Walker}, {Nadeau}, {Doyon}, \&
  {Marois}}]{Racine1999}
{Racine}, R., {Walker}, G.~A.~H., {Nadeau}, D., {Doyon}, R., \& {Marois}, C.
  1999, \pasp, 111, 587

\bibitem[{{Samland} {et~al.}(2021){Samland}, {Bouwman}, {Hogg}, {Brandner},
  {Henning}, \& {Janson}}]{Samland2021}
{Samland}, M., {Bouwman}, J., {Hogg}, D.~W., {et~al.} 2021, \aap, 646, A24

\bibitem[{{Samland} {et~al.}(2017){Samland}, {Molli{\`e}re}, {Bonnefoy},
  {Maire}, {Cantalloube}, {Cheetham}, {Mesa}, {Gratton}, {Biller}, {Wahhaj},
  {Bouwman}, {Brandner}, {Melnick}, {Carson}, {Janson}, {Henning}, {Homeier},
  {Mordasini}, {Langlois}, {Quanz}, {van Boekel}, {Zurlo}, {Schlieder},
  {Avenhaus}, {Beuzit}, {Boccaletti}, {Bonavita}, {Chauvin}, {Claudi}, {Cudel},
  {Desidera}, {Feldt}, {Fusco}, {Galicher}, {Kopytova}, {Lagrange}, {Le
  Coroller}, {Martinez}, {Moeller-Nilsson}, {Mouillet}, {Mugnier}, {Perrot},
  {Sevin}, {Sissa}, {Vigan}, \& {Weber}}]{Samland2017}
{Samland}, M., {Molli{\`e}re}, P., {Bonnefoy}, M., {et~al.} 2017, \aap, 603,
  A57

\bibitem[{{Sauvage} {et~al.}(2016){Sauvage}, {Fusco}, {Petit}, {Costille},
  {Mouillet}, {Beuzit}, {Dohlen}, {Kasper}, {Suarez}, {Soenke}, {Baruffolo},
  {Salasnich}, {Rochat}, {Fedrigo}, {Baudoz}, {Hugot}, {Sevin}, {Perret},
  {Wildi}, {Downing}, {Feautrier}, {Puget}, {Vigan}, {O'Neal}, {Girard},
  {Mawet}, {Schmid}, \& {Roelfsema}}]{Sauvage2016}
{Sauvage}, J.-F., {Fusco}, T., {Petit}, C., {et~al.} 2016, Journal of
  Astronomical Telescopes, Instruments, and Systems, 2, 025003

\bibitem[{{Schmid} {et~al.}(2018){Schmid}, {Bazzon}, {Roelfsema}, {Mouillet},
  {Milli}, {Menard}, {Gisler}, {Hunziker}, {Pragt}, {Dominik}, {Boccaletti},
  {Ginski}, {Abe}, {Antoniucci}, {Avenhaus}, {Baruffolo}, {Baudoz}, {Beuzit},
  {Carbillet}, {Chauvin}, {Claudi}, {Costille}, {Daban}, {de Haan}, {Desidera},
  {Dohlen}, {Downing}, {Elswijk}, {Engler}, {Feldt}, {Fusco}, {Girard},
  {Gratton}, {Hanenburg}, {Henning}, {Hubin}, {Joos}, {Kasper}, {Keller},
  {Langlois}, {Lagadec}, {Martinez}, {Mulder}, {Pavlov}, {Podio}, {Puget},
  {Quanz}, {Rigal}, {Salasnich}, {Sauvage}, {Schuil}, {Siebenmorgen}, {Sissa},
  {Snik}, {Suarez}, {Thalmann}, {Turatto}, {Udry}, {van Duin}, {van Holstein},
  {Vigan}, \& {Wildi}}]{Schmid2018}
{Schmid}, H.~M., {Bazzon}, A., {Roelfsema}, R., {et~al.} 2018, \aap, 619, A9

\bibitem[{{Skemer} {et~al.}(2015){Skemer}, {Hinz}, {Montoya}, {Skrutskie},
  {Leisenring}, {Durney}, {Woodward}, {Wilson}, {Nelson}, {Bailey}, {Defrere},
  \& {Stone}}]{Skemer2015}
{Skemer}, A.~J., {Hinz}, P., {Montoya}, M., {et~al.} 2015, in Society of
  Photo-Optical Instrumentation Engineers (SPIE) Conference Series, Vol. 9605,
  Techniques and Instrumentation for Detection of Exoplanets VII, ed.
  S.~{Shaklan}, 96051D

\bibitem[{{Snik} {et~al.}(2012){Snik}, {Otten}, {Kenworthy}, {Miskiewicz},
  {Escuti}, {Packham}, \& {Codona}}]{Snik2012}
{Snik}, F., {Otten}, G., {Kenworthy}, M., {et~al.} 2012, in Society of
  Photo-Optical Instrumentation Engineers (SPIE) Conference Series, Vol. 8450,
  Modern Technologies in Space- and Ground-based Telescopes and Instrumentation
  II, ed. R.~{Navarro}, C.~R. {Cunningham}, \& E.~{Prieto}, 84500M

\bibitem[{{Soummer}(2005)}]{Soummer2005}
{Soummer}, R. 2005, \apjl, 618, L161

\bibitem[{{Sparks} \& {Ford}(2002)}]{Sparks2002}
{Sparks}, W.~B. \& {Ford}, H.~C. 2002, \apj, 578, 543

\bibitem[{{Vigan}(2020)}]{Vigan2020_pipeline}
{Vigan}, A. 2020, {vlt-sphere: Automatic VLT/SPHERE data reduction and
  analysis}, Astrophysics Source Code Library, record ascl:2009.002

\bibitem[{{Vigan} {et~al.}(2021){Vigan}, {Fontanive}, {Meyer}, {Biller},
  {Bonavita}, {Feldt}, {Desidera}, {Marleau}, {Emsenhuber}, {Galicher}, {Rice},
  {Forgan}, {Mordasini}, {Gratton}, {Le Coroller}, {Maire}, {Cantalloube},
  {Chauvin}, {Cheetham}, {Hagelberg}, {Lagrange}, {Langlois}, {Bonnefoy},
  {Beuzit}, {Boccaletti}, {D'Orazi}, {Delorme}, {Dominik}, {Henning}, {Janson},
  {Lagadec}, {Lazzoni}, {Ligi}, {Menard}, {Mesa}, {Messina}, {Moutou},
  {M{\"u}ller}, {Perrot}, {Samland}, {Schmid}, {Schmidt}, {Sissa}, {Turatto},
  {Udry}, {Zurlo}, {Abe}, {Antichi}, {Asensio-Torres}, {Baruffolo}, {Baudoz},
  {Baudrand}, {Bazzon}, {Blanchard}, {Bohn}, {Brown Sevilla}, {Carbillet},
  {Carle}, {Cascone}, {Charton}, {Claudi}, {Costille}, {De Caprio},
  {Delboulb{\'e}}, {Dohlen}, {Engler}, {Fantinel}, {Feautrier}, {Fusco},
  {Gigan}, {Girard}, {Giro}, {Gisler}, {Gluck}, {Gry}, {Hubin}, {Hugot},
  {Jaquet}, {Kasper}, {Le Mignant}, {Llored}, {Madec}, {Magnard}, {Martinez},
  {Maurel}, {M{\"o}ller-Nilsson}, {Mouillet}, {Moulin}, {Orign{\'e}}, {Pavlov},
  {Perret}, {Petit}, {Pragt}, {Puget}, {Rabou}, {Ramos}, {Rickman}, {Rigal},
  {Rochat}, {Roelfsema}, {Rousset}, {Roux}, {Salasnich}, {Sauvage}, {Sevin},
  {Soenke}, {Stadler}, {Suarez}, {Wahhaj}, {Weber}, \&
  {Wildi}}]{Vigan2021_shine}
{Vigan}, A., {Fontanive}, C., {Meyer}, M., {et~al.} 2021, \aap, 651, A72

\bibitem[{{Wang} {et~al.}(2015){Wang}, {Ruffio}, {De Rosa}, {Aguilar}, {Wolff},
  \& {Pueyo}}]{pyklip2015}
{Wang}, J.~J., {Ruffio}, J.-B., {De Rosa}, R.~J., {et~al.} 2015, {pyKLIP: PSF
  Subtraction for Exoplanets and Disks}, Astrophysics Source Code Library,
  record ascl:1506.001

\end{thebibliography}


\end{document}